\documentclass[preprint,12pt]{elsarticle}
\usepackage{multirow}
\usepackage{listings}
\usepackage{subcaption}
\usepackage[T1]{fontenc}
\usepackage[switch]{lineno}
\usepackage{caption}
\usepackage[hidelinks]{hyperref}
\usepackage{colortbl}
\usepackage{xcolor}
\usepackage{threeparttable}
\usepackage{tikz}
\usepackage{enumitem}
\usepackage{pgfplots}
\usepackage{filecontents}
\usepackage{datatool}
\usepackage{afterpage}
\usepackage{tkz-fct}
\usepackage{footnote}
\usepackage{makecell}
\usepackage{rotating}
\usepackage{fontawesome}
\usepackage{amssymb}
\usepackage{MnSymbol,wasysym}

\usepackage{booktabs}
\usepackage{calc}

\pgfplotsset{compat=1.17, legend cell align=left}

\definecolor{female}{HTML}{FDAE61}
\definecolor{male}{HTML}{536872}
\definecolor{lightgray}{rgb}{0.83, 0.83, 0.83}
\definecolor{silencecolor}{rgb}{0.0, 0.18, 0.39}

\newlength\MAX  

\newlength\WIDTHOFBAR
\newcommand*\femalewhitebar[1]{#1~\rlap{\textcolor{female!30}{\rule{\WIDTHOFBAR}{2ex}}}\textcolor{female!100}{\rule{#1\WIDTHOFBAR}{2ex}}}

\def\showblankfemalewhitebar#1{
  {\color{female!100}\rule{#1cm}{6pt}}{\color{female!100}\rule{\WIDTHOFBAR - #1 cm}{6pt}}}
\newcommand*\malewhitebar[1]{#1~\rlap{\textcolor{male!30}{\rule{\WIDTHOFBAR}{2ex}}}\textcolor{male!100}{\rule{#1\WIDTHOFBAR}{2ex}}}

\def\showblankmalewhitebar#1{
  {\color{male!100}\rule{#1cm}{6pt}}{\color{male!100}\rule{\WIDTHOFBAR - #1 cm}{6pt}}}

\begin{document}
\begin{sloppy}
\title{Gender Influence on Student Teams' Online Communication in Software Engineering Education}
\journal{The Journal of Systems and Software}

\author[1]{Rita Garcia}
\author[2]{Christoph Treude}

\address[1]{Victoria University of Wellington, Wellington, New Zealand}
\address[2]{Singapore Management University, Singapore}

\begin{abstract}
Collaboration is crucial in Software Engineering (SE), yet factors like gender bias can shape team dynamics and behaviours. This study examines an eight-week project involving 39 SE students across eight teams contributing to GitHub projects. Using a mixed-methods approach, we analysed Slack communications to identify gender differences, comparing how they influence learning gains. We found higher help-seeking and leadership behaviours in the all-woman team, while men responded more slowly. Although communication did not affect final grades, we identified statistical significance correlating communications with students' understanding of software development. With some students putting more effort into collaboration, future work can investigate diversity and inclusion training to balance these efforts. The observed link between team engagement and a higher understanding of software development highlights the potential for teaching strategies that promote help-seeking. These findings could guide efforts to address challenges student SE teams face when using communication platforms and foster more equitable collaborative learning in Software Engineering Education.
\end{abstract}


\maketitle

\textbf{Author's Note:} This paper extends our previous six-page conference paper through additional analysis and expanding the work to answer three instead of one research questions \citep{garcia:2021}.

\section{Introduction}

Software Engineering (SE) is a collaborative process that involves people working together to create software applications and services \citep{whitehead:2007}. Collaboration involves conflict resolution, decision-making, problem-solving, and communication skills \citep{webb:1995}. Poor communication might arise when students collaborate on assessments through teamwork, potentially influencing how they work together \citep{vanhanen:2018}. Poor communication could result from biases that consciously or unconsciously influence a student's behaviour \citep{wang:2019}. For example, \textit{gender bias}, the preferential treatment towards one gender over another \citep{moss:2012}, can contribute to the barriers that women experience while collaborating on software engineering projects \citep{lee:2019}. 

Through the lens of gender, we performed a study that examined the gender differences emerging when student SE teams contribute to a large open-source software project. This paper builds on previous findings \citep{garcia:2021} that examined through gender analysis how student teams initiated online communication during collaborative Software Engineering Education. The study observed significant statistical results with how the women participating initiated communication. The women in the student groups applied more team leadership, project coordination, and monitoring necessary to complete the project. The prior study found that members sought more help from peers within the all-women team, an infrequent behaviour within the all-man and mixed-gender teams. Because the conference proceedings limited our original paper to six pages, we use this paper to examine how the student team members respond to the initiated communications, providing an overview of how student teams communicate online during the software collaboration process. In this paper, we can investigate further how the collaboration process and students' communication on the project might have influenced their learning gains in the course. We evaluate the interdependence between team communications and learning gains because improved communication skills foster a connection between proactive collaboration and documented advancements in students' problem-solving \citep{terenzini:2001}. To the best of our knowledge, this is the first study that examines students' communications in-depth while completing a software engineering project. Our study focuses on communication through the lens of gender because it is a notable social category and is a starting point to examine students' collaborative collaboration. Our work provides future opportunities to re-examine or replicate this study that considers other social categories and intersectionality, ``a unique experience that is separate and apart from its originating categories'' \citep[p.~3]{rosette:2018}. This paper presents the gender analysis on students' online communications, answering the following research questions:

\begin{itemize}
   \item \textit{RQ1: How does gender influence students' initiation of communication within teams?}
   \item \textit{RQ2: How does gender influence students' responses to the initiated communications?}
   \item \textit{RQ3: How do the communication behaviours between women and men in collaborative team settings impact their learning gains?}
\end{itemize}

To answer these questions, we examined the communications of 39 SE students working in eight teams of 4-6 people communicating over Slack \citep{slack:2020}, an online messaging platform. We share our observations of the SE student teams working together to contribute to an open-source software project. Though these are communication behaviours from one SE course, our findings might help practitioners realise the difference in communication behaviours, potentially resulting in different roles and responsibilities for the women and men collaborating on the projects. We use a mixed-methods approach to analyse the data collected from an eight-week project. The study used gender analysis ``to assess differences in participation, benefits and impacts between men and women, including progress towards gender equality and changes in gender relations'' \citep[p.~100]{hunt:2004}. Pre-post surveys were used to collect students' perceptions of teamwork; we then analysed their online communications to identify their teamwork behaviours. The analysis examined teams' communications and how members reacted to the initiated communications. The results also showed statistical significance in students' learning gains with an individual quiz. In addition, we observed higher marks in a self-reflection activity for students contributing to group discussions. Team communication might have given these students a deeper understanding to complete these activities with higher marks. The teams' responses to initiated communications show the women following up on tasks while the men shifted the conversation to other topics. Our findings raise future research opportunities to investigate further and mitigate inequalities in the teamwork we observed in this study performed in an SE course. Our contributions could assist Software Engineering Education researchers and practitioners in foreseeing obstacles students might encounter collaborating on software projects using communication platforms in comparable learning environments to our research.

\section{Background} \label{section_background}

Through the lens of gender, our work examines student teams' online communications that occur during collaborative learning in Software Engineering Education. With SE student teams working on a software development project, conflicts might emerge during team communications that impede learning \citep{johnson:1979}. Our interest is to examine the communications and compare the approaches women and men use during collaboration and their influence on learning gains. We are interested in observing negative or inequitable communications that may influence women's waning interests in computing \citep{margolis:2000}. In this section, we review existing literature that supports our work and helps us better understand how our research is situated within the existing literature. We divide our review into three areas. First, Section \ref{section_gender_theory} presents a broad background on gender-related social behaviours, gender bias, and gender differences during communication between women and men. This section examines the experiences that may influence women during collaborative learning and uses surveys and textual analysis to report on observed gender differences in behaviours, interactions, and interpersonal communications between women and men. Secondly, Section \ref{section_retention} reviews literature that suggests factors that influence women leaving computing degrees early and the strategies educators can take to encourage the retention of women. Lastly, since our work focuses on SE students' online communication, Section \ref{section_collab_learning} provides background on communication exchange that occurs during collaborative learning. We use this background to strengthen and explain the communication structure and exchange observed in our study. We build on the prior literature we present in Sections \ref{section_gender_theory}-\ref{section_collab_learning} to continue the examination of the different communication behaviours that women and men employ. Our work examines whether these differences influence inequalities in the student's participation in the project, including any influence it may have on their learning gains in the Software Engineering course.

\vspace*{5mm}\subsection{Gender-Related Behaviours, Interactions, and Interpersonal Communication} \label{section_gender_theory}

This section presents literature on gender-related social behaviours, gender bias, and gender differences that emerge when women and men communicate. The literature gives us background on potential behaviours we may observe during our research. When trying to understand gendered interactions, it is essential to consider how society attaches inequality to gender. ``Gender is a system of social practices within society that constitutes people as different in socially significant ways and organizes relations of inequality on the basis of the difference'' \citep[p.~247]{ridgeway:1999}. According to \citet{ridgeway:1999}, the confirmation of the gender system involves the cultural acceptance that women and men are different, reinforcing gendered inequalities. As a result, the system of social practices shapes gender as a social structure containing rules and schemas defined by culture for women and men \citep{sewell:1992}. These rules and schemas influence the roles and norms of women and men at different layers of society. The social structure of the gendered rules and schemas is defined within three dimensions to form a conceptual framework \citep{risman:2013}. The first dimension, \textit{Individual}, focuses on the progress of gender identity and how the individual socialises. The second dimension, \textit{Interactional}, focuses on the cultural expectations for women and men and the presumptions placed on them even when performing equivalent societal roles. The third and final dimension, \textit{Institutional}, examines how culture influences actions through regulations and gender-specific organisational schemas. The creators of the conceptual framework, \citet{risman:2013}, posit that the dimensions have a cyclic relationship,  influencing the dynamics in the social structure that can contribute to society's gender inequalities. 

Because inequalities are drawn from the different social structure dimensions, researching and discussing social interactions can be complex. To better understand and investigate social behaviours during gender interactions, a model \citep{deaux:1987} was constructed to represent these interactions. This interactive model considers the conditions that impact how gender-related behaviours occur during interactions. The interactive model has three components. The first component is the \textit{Perceiver}, an individual that brings to the social interaction their gender beliefs and their personal goals for the exchange. The second component is the \textit{Target}, an individual that brings to the social interaction their individual gender-related beliefs about themselves, including their abilities, behaviours, and characteristics. The target also brings their goals to the exchange. The last component is the \textit{Situation}, a circumstance that influences the degree of gender-related issues in the interaction. The situation can affect how the perceiver and target present themselves in public or private spaces.

The interactive model can help examine \textit{interpersonal communication}, where perceivers and targets exchange ideas, information, and feelings about a given context \citep{whiteman:2002}. Prior research has investigated the approach and behaviours women and men use during interpersonal communication. For example, one study \citep{popp:2003} generalises how the genders' communicate, characterising women's communication as more emotional and less direct and men's communication approach as more authoritarian, forceful, and blunt. Though these characteristics can be considered stereotypes, early research \citep{scott:1980} on gender communication views women and men as having few coinciding characteristics that can affect communication between women and men. A review \citep{troemel:1991} of Tannen's \textit{You Just Don't Understand} \citep{tannen:1990}, a book that discusses conversational styles of women and men, analyses the presented dialogues between the genders, raising generalisations of the communication styles. The review discusses how women use communication to build trust and relationships in their teams. Men use communication to assert dominance and use their connections to achieve their goals. Though Tannen claims the two conversation styles are ``equally valid'', the analysis showed how the men's approach in face-to-face conversations would dominate while the women yielded, demonstrating how the social construct between women and men influenced their conversations. When virtual teams use asynchronous communication platforms to collaborate, another study \citep{aries:1996} observed that some face-to-face communication strategies, like verbal interruptions, are not as effective or plausible. Some ineffective behaviours, like verbal interruptions, are absent in virtual team environments \citep{furumo:2007}. As a result, women collaborating in virtual team environments have a higher satisfaction rate than men since ineffective behaviours are not as potent when applied in asynchronous communication interactions.

In this section, we reviewed literature that demonstrates the complex social factors between women and men that affect gender-related behaviours, interactions, and interpersonal communications. However, other factors beyond gender can influence women's and men's behaviours, interactions, and communications. These factors include social categories, such as social identity, race, and past circumstances and relationships. The overlapping social categories, or intersectionality, suggest scholarship should consider these categories when evaluating data. Unfortunately, the different social structures of inequality, such as race and gender, as previously discussed, do not follow the same arrangement, requiring more effort on how the social structures intersect during social behaviours \citep{risman:2004}. As a result, sometimes, in gender research, conclusions are drawn on gender without consideration of the other social layers. For example, a meta-analysis study by \citet{paustian:2014} reviewed the literature on gender differences in perceptions of leadership effectiveness. The meta-analysis uses a theoretic framework, role congruity theory (RCT), as a basis for the study. RCT focuses on the discrimination and biases women encounter in leadership roles \citep{eagly:2002}. Focusing on the single social structure allowed the authors to compare their findings with other meta-analysis studies on gender differences. 

Additionally, further research \citep{popp:2003} in the field of communication recognises intersectionality as not explored when forming results. However, \citet{popp:2003} acknowledge existing research focusing on one social structure, gender or race. The findings in the literature presented in the section raise awareness of observed challenges women experience during collaborative learning, and the literature suggests strategies to minimse these experiences. For our study, we begin our evaluation of interpersonal communications through the lens of gender since it is a notable social category, allowing us to re-examine or replicate the study for nuanced results to consider other social categorisations and intersectionality. 

\subsection{Retention of Women in Computing} \label{section_retention}

It is well established that women are under-represented in the computing disciplines. Statistics from 2016 show that 19\% of the CS Bachelor's degrees were awarded to women in the United States, a recorded decline from the 27\% awarded in 1997 \citep{nces:2016}. Some educators perceive the declining numbers of women in computing may be due to misinformation among young women about the rewards of a career in computing \citep{salminen:2011}. Other reasons may be due to women's learning experiences in computing, which we discuss later in this section. 

The computing fields have examined retention strategies to increase women's representation \citep{cohoon:2002}. Retention strategies are necessary in education because women sometimes change their computing majors after their first year due to waning interests \citep{margolis:2000}. Changing majors is sometimes the result of low self-confidence, where they perceive men as better in computing \citep{margolis:2000}. This gender difference in self-belief impacts an individual's self-confidence, creating a \textit{confidence gap} that affects a woman's ability to attempt tasks and generating a fear of failure that can impact their achievement and learning, such as mathematics \citep{ross:2012}. The lower self-confidence in abilities can also stem from a \textit{sense of belonging}, ``a person's experience of being valued or important to an external referent and experiencing a fit between self and that referent'' \citep[p.~174]{hagerty:1992}. Women's sense of belonging in the field of computing might be influenced by their perception their interests and focus in computing are not as strong as the men \citep{margolis:2002}. Women's perceptions and self-confidence can come from stereotypes that nerds and hackers are better suited for computing careers \citep{cuny:2002}, which can detract women from pursuing a degree in these fields. Instead, women's interest in computing applies to other sectors, such as the sciences. In contrast, men gravitate to computing for gaming or have been an activity encouraged by outside influencers, such as parents, through access to computers at a young age. To promote a sense of belonging and reduce the confidence gap in computing, faculty can get involved by adjusting the classroom environment to make it more inclusive to women \citep{cheryan:2011} and have faculty role models to help them feel less isolated within large courses \citep{cuny:2002} and peer support to build confidence, which plays a critical role in women majoring in a computing course \citep{margolis:2000}, such as Software Engineering.

For this paper, we reviewed literature that examined women's learning experiences in computing courses. We wanted to understand better women's motivations for pursuing other majors. The first study we examined was conducted by \citet{cox:2008}. In this study, the researchers surveyed women studying Software Engineering (SE) within an all-woman team and compared their prior experiences collaborating in mixed-gender teams. The results showed higher levels of cooperation in the all-woman team, encouraging each other to attempt new tasks. However, within the mixed-gender teams, one woman described the collaborative learning environment as non-inclusive, noting that the men ``don't allow the girl(s) in a group to fully participate'' \citep[p.~8]{cox:2008}. In another study by \citet{wang:2019}, \textit{intergroup contact theory} was applied to \textit{collaborative learning}, an approach with students working together to understand learning concepts better \citep{webb:1995} to reduce implicit gender bias. Intergroup contact theory attempts to counteract bias by bringing different social groups, such as women and men, together for interpersonal interactions \citep{pettigrew:1998}. In this study, 280 SE students formed 70 mixed-gender teams to collaborate on an eight-week software project. To evaluate whether intergroup contact theory had any influence, participants were given pre-post assessments to measure changes in their implicit gender bias. The results showed a reduction in men's implicit gender bias when they collaborated in a predominately woman mixed-gender team. However, these changes were not apparent in women's implicit gender bias while collaborating in a mixed-gender team predominately made up of men.

We also examined activities that occur during collaboration, evaluating how they influence women's learning experiences during collaboration. During the learning process, activities such as student questioning can allow students to learn from peers \citep{boud:1999} and can help reduce misconceptions \citep{collins:1985}. However, social barriers sometimes hinder students from asking questions because they do not want to appear ``ignorant'' \citep{graesser:1994} to their peers. A study by \citet{sankar:2015} evaluated the questions and answers of more than a million STEM students on a Q\&A forum, Piazza, to examine women's feelings of isolation during the learning process. Women were less involved in answering questions and used Piazza's anonymity feature more than the men. The results suggest women might feel more confident contributing to questioning activities, but only anonymously, which helps them feel less isolated in their learning. A later Piazza study \citep{thinnyun:2021} evaluated gender differences in students' engagement on the forum. This study collected data from 2500 Piazza users, analysing the questioning frequency and length of engagement. The results confirmed that women use the anonymity feature more, which enables them to ask more questions and spend more time on the forum. These studies suggest women have apprehension about contributing during collaborative learning, but strategies such as anonymity could help them feel less isolated and give them the confidence to contribute. 

Throughout this section, we present literature within Computing Education Research that tries to understand why women leave computing degrees early and strategies educators can take to retain them. This research community could leverage the efforts from gender studies to evaluate different approaches or strategies to boost women's representation in the field of computing \citep{salminen:2011}. Engaging gender studies in the context of computing education can help identify effective methods and areas needing improvement in initiatives geared towards increasing and retaining women in computing.

\subsection{Communication Exchange During Collaborative Learning} \label{section_collab_learning}

The literature presented in Section \ref{section_retention} shows interpersonal interactions potentially influencing women's learning experiences. In this section, we examine the communication structure to understand better how individuals use it to exchange information during \textit{collaborative learning}. Collaborative learning is the underlying educational theory and pedagogy educators use to bring learners together to master concepts \citep{cohen:1994}. This section relates to Section \ref{section_retention} because interpersonal interactions discussed in the previous section are used during collaborative learning, and collaborative learning helps students develop teamwork skills for professional careers \citep{oneil:1992}. Teamwork skills require guidance and training \citep{benne:1948} to help the team members understand the essential roles necessary for facilitating growth and productivity within the team. In the absence of guidance, sometimes individual personalities have been confused with the assumed team roles, which can exclude students from fully participating or practising roles and responsibilities in the team and exclude them from fully participating in the collaborative learning process. In addition, challenges during the collaboration process sometimes manifest due to social pressures that might lead to conflicts that can interfere with learning \citep{johnson:1979}. Social pressures could be gender-related, resulting in the inequitable division of labour within mixed-gender teams \citep{deaux:1987}. For example, gender-related social pressures may hinder women from being \textit{initiator-contributors} who recommend new ideas or alternative perspectives to the team's goals \citep{benne:1948}.

During collaborative learning, students find ``information themselves, evaluating and critically analysing the information, discussing it with one another, building structured arguments and drawing conclusions about various topics under discussion'' \citep[p.~169--170]{ellis:2001}. In face-to-face collaborative learning, oral communication is a strategy for knowledge exchange. With oral communication, \textit{speakers} vocalise messages, while \textit{hearers} are the recipients of that message \citep{sacks:1974}. In computer-mediated communications (CMC), the speaker becomes the \textit{writer}, and the hearers become the \textit{recipients} \citep{garcia:1999}. During CMC, communication exchange is supported by technologies such as networked computers, social network sites, and message services \citep{yao:2020}. \citet{sacks:1974} explained communication exchange between speakers and hearers (writers and recipients) is done through \textit{turn-taking}. Turn-taking occurs when a participant becomes the speaker (writer) in communications. There are three components in a turn-taking structure. We describe these components in CMC environments:
\begin{enumerate}
    \item \textit{Turn-Constructional Component:} A writer's turn contributing to a conversation. 
    \item \textit{Turn-Allocation Component:} Techniques that help determine the next writer, either the writer selecting the next person or deciding through self-selection.
    \item \textit{Rules:} Rules govern the turn-taking structure. These rules order the turn-taking until the conversation ends with no other speaker self-selecting to contribute to the conversation or the current speaker beginning another turn-taking component, which changes the conversation focus to another topic.
\end{enumerate}

In addition to defining the turn-taking structure, \citet{sacks:1974} also identified 14 conversation behaviours and attributes present during communication exchange, including the number of speakers, the type of talk (continuous or discontinuous), and the type of turn-constructional units. These behaviours and attributes help model turn-taking in conversations.

Turn-taking has been previously used to examine Socratic dialogues, where tutors pose questions rather than explanations to support students' learning \citep{core:2003}. In the \citet{core:2003} study, turn-taking was classified to evaluate the interactive nature of the Socratic dialogues. These dialogue sessions were more interactive than didactic dialogues due to the tutor's initiative to prompt students to encourage communication. Turn-taking has also been used to evaluate students' motivation during one-on-one tutoring sessions for introductory Computer Science courses between a student and tutor \citep{boyer:2009}. The study by \citet{boyer:2009} used turns to structure the dialogues with the students' problem-solving actions, labeling turns as student and tutor-initiative that denotes the person controlling the problem-solving effort. By analysing the turn-taking in the dialogues, the study observed the instructional strategies during these sessions can influence motivational outcomes, such as increased self-confidence to less-confident students receiving praise or reassurance during the sessions. By analysing turn-taking, researchers can examine dialogues in-depth, helping to understand better the motivations behind communications initiated by the learner. 

\begin{figure}[h!]
\centering
  \includegraphics[width=0.7\textwidth]{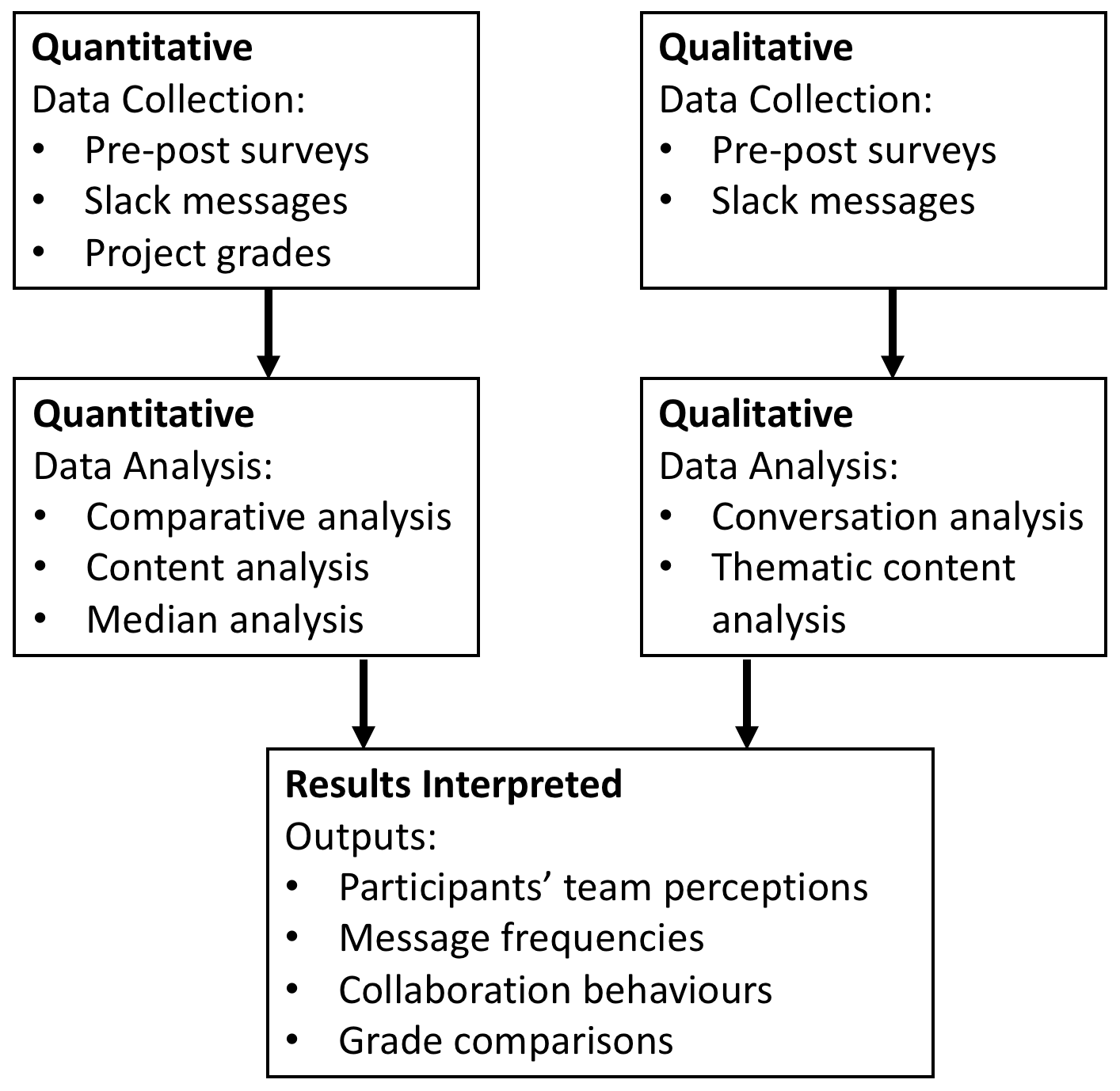}
    \caption{Diagram of Study Method (Adapted from \citet{creswell:2006})}
  \label{fig:study_method_figure}
\end{figure}

\section{Methods}

We adopted a mixed-methods approach \citep{creswell:2012} using a triangulation design \citep{creswell:2006} that interprets the collected data. Figure \ref{fig:study_method_figure} shows the parallel consideration given to the quantitative and qualitative data sources. The quantitative method analyses the data from the pre-post surveys and the Slack messages, while the qualitative method provides an in-depth view of students' communications. In this section, we present the study's context (Section \ref{subsection_context}) and the pre-post surveys (Section \ref{subsection_prepost_surveys}) used in this study. This section explains the quantitative and qualitative analyses performed on the collected data. Section \ref{data_quant_analysis} describes the quantitative analysis, while Section \ref{data_qual_analysis} presents the qualitative analysis performed on the collected data. The quantitative and qualitative analysis sections contain information to construct the findings on student teams' initiated communications. This information was previously reported \citep{garcia:2021}, but we include it in this paper as background and context for how students responded to initiated communications.

\subsection{Context} \label{subsection_context}

The study was conducted in a 12-week SE course (Semester 1 2021) at a large university, teaching best practices in software development. The authors received a grant to perform the study. The course enrolled 79 students, 14 (18.00\%) women and 65 (82.00\%) men. Based on the course roster, none of the enrolled students identified as non-binary. The enrolled students were under- (55.70\%, n=44) and postgraduate (44.30\%, n=35) students. We bring in students' under- (UG) and postgraduate (PG) status to better understand the study's context and giving higher confidence about the learning settings to which our findings can be applied. To reinforce learning objectives, this course had students working in teams to contribute to a large open-source software project hosted on GitHub. The lecturer encouraged students to form teams. The lecturer guided the project's goals and requirements but did not assign team roles for completing the work. For their projects, the course lecturer encouraged teams to select one of three projects external to the university:

\begin{enumerate}
    \item \textit{JabRef} \citep{jabref:2021}: A literature management system used to organise references.
    \item \textit{Pygments} \citep{pygments:2021}: A Python-based syntax highlighter for multiple platforms, such as programming languages and wikis.
    \item \textit{Zettlr} \citep{zettlr:2021}: A markdown editor designed for a broad range of users, including researchers, students, and authors.
\end{enumerate}

The course used these three open-source projects because they support developers, especially students new to the open-source community. These three projects contain detailed instructions to help guide novices who are contributing to them. The projects have similar practices for the team to contribute. The course gave students three project options since one project could only be able to handle some of the student teams. For example, the project maintainers\textemdash~the core project team\textemdash~might not be able to answer all the teams' questions in a timely fashion. In addition, the enrolled students have different programming language backgrounds. The course lecturer allowed the students to select a project using a language they were comfortable with, making it equitable for all students.

Teams used a pull-based software development model to contribute software changes to be considered by the project maintainers \citep{gousios:2014}. The software changes are packaged as \textit{Pull Requests} (PRs), suggested changes for project maintainers to review and comment on before accepting \citep{kononenko:2018}. The project maintainers provide feedback on the teams' PRs on the GitHub platform, where the teams use a git feature branch as a shared space to contribute to the project. The teams communicated with the project maintainers on the platform to improve their software changes. Upon acceptance, PRs are integrated into the GitHub source repository. 

The lecturer notified the project maintainers for the three projects in advance of the course, letting them know students would be contributing to their projects, submitting PRs, and asking questions. The lecturer asked the project maintainers to respond to the students promptly and politely. Teams selected open issues from the GitHub projects that interested them. The lecturer made the final decision on one issue for each team, coordinating the issues to ensure teams' contributions did not overlap. The teams were required to submit at least one PR to the GitHub project. 

\begin{table}[]
\centering
\fontsize{9pt}{9pt}\selectfont
\caption{Course Assessments used in the Study}\label{table_course_assessment}
\setlength\extrarowheight{2pt}
\begin{tabular}{p{2.8cm}|c|c|c|p{5cm}}
\textbf{Assessment} & \textbf{Weight} & \textbf{Mode} & \textbf{Type} &  \textbf{Description}  \\ \hline
Contribution \newline analysis & 20\% & Individual & Formative & A two-page essay where students analyse the contribution processes of two successful open source contributors.\\ \hline
Contribution \newline proposal & 10\% & Team & Formative & An early team presentation, where teams identified their project and plans for contributing to the open-source project.   \\ \hline
Open-source \newline contribution & 20\%  & Team & Summative & A presentation conducted at the end of the project, presenting how the teams contributed to the open-source project. Outcomes and challenges were discussed, explaining what was learned from the process.   \\ \hline
Quiz & 10\% & Individual & Summative & An assessment that reinforces students' understanding about contributing to open-source projects. \\ \hline
Reflection \newline essay & 20\% & Individual & \small{Summative} & A two-page essay, where students reflect on their impressions before and after contributing to the open-source project. \\ \hline
Team \newline communication & 20\% & Individual & Formative & Students' communications conducted over Slack. \\ 
\end{tabular}
\end{table}

The course contained six assessments, shown in Table \ref{table_course_assessment}. The table lists the assessments in alphabetical order with a brief description and their contribution to the final grade for the course. Two assessments, \textit{Contribution proposal} and \textit{Open-source contribution}, are collaborative learning activities. The \textit{Open-source contribution} relates to the teams' PRs. Teams were required to submit the URLs for the PR completion of the project. Teams received the same marks for the two collaborative activities. The remaining four assessments, \textit{Analysis of existing contribution process}, \textit{Quiz}, \textit{Reflection essay}, and \textit{Team communication}, are independent activities. Student teams were given eight weeks to complete the project, with two deliverables: a project proposal and a final presentation. The entire team gave the final presentations in week eight. Teams shared their results and experiences collaborating on the open-source project during the presentations. The marking criteria for under- and postgraduate students were the same.

The \textit{Team communication} assessment tracked the students' online communications. The individual's contribution to the communication was worth 20\% of their course grade. The authors discussed different tools that enabled students to discuss the project while allowing them to use collaboration tools commonly applied in professional software development. Though a prior study \citep{seppala:2016} showed students have experience using communication and collaboration tools, such as GitHub and Google Chat, the authors selected Slack \citep{slack:2020} for online communication. Slack gives students real-world experiences using tools that support professional software development \citep{lin:2016} while giving the authors administrative rights to monitor students and collect data from the communication platform. Slack channels were created for each team on the university's workspace, a meta-container that stores and manages channels. Channels are private spaces for teams to communicate. Each channel was labelled with the team IDs, such as Team-1. The lecturer instructed students to use their Slack channels for team collaboration, providing guidance that included the use of polite and responsive communication for effective collaboration. Since the lecturer marked the \textit{Team communication} assessment from the Slack messages, the lecturer encouraged them to use the teams' Slack channels for all their communications. The lecturer also explained that the Slack messages should be meaningful so that someone outside their team could comprehend their actions through the messages. To evaluate \textit{Team communication}, the course lecturer examined the frequency and quality of the students' Slack messages using a Slack administration report that helped determine how active individuals were on the platform. 

Students from the course were invited to participate in the study. Ethical approval was obtained from the university's ethics committee to conduct the study. To recruit volunteers, an announcement was made by the main author during a class session and posted on Canvas, the Learning Management System (LMS) is used by the university to administer the course and instructional materials. Students who participated in the study received a USD\$60 voucher to encourage a higher participation rate from the enrolled students. The recruitment announcement explained that participants would receive a voucher but did not specify the amount. After we recruited volunteers, we determined the voucher amount by evenly distributing the remaining grant as vouchers. 

\subsection{Pre-Post Surveys} \label{subsection_prepost_surveys}

The pre-post surveys were designed to collect participants' teamwork experiences.\footnote{https://doi.org/10.6084/m9.figshare.15026037} The surveys were constructed from previous instruments that measured gender bias \citep{diehl:2020}, gender inequality \citep{popp:2019}, and team collaboration \citep{tafliovich:2016}. We adjusted the questions from these instruments to frame the questions from the student's perspective. 

The pre-survey was designed to collect participants' perceptions of teamwork through prior experiences. The 14-question pre-survey was also designed to determine participants' previous gender bias and inequity experiences. Half (n=7) of the pre-survey questions were refactored from a gender bias study \citep{diehl:2020}. The questions were derived from a framework \citep{diehl:2016} that identified distinct barriers that cause gender bias. The barriers include male privilege, disproportionate constraints, insufficient support, devaluation, and hostility. By setting barrier-based questions into the pre-survey, a deeper understanding of the gender bias factors that participants face in the learning environment was gained, pinpointing any positive or negative presumptions they have about gender bias.

To complete the pre-survey, participants were required to provide their full name and gender. Gender options included \textit{Women}, \textit{Men}, \textit{Prefer not to say}, and \textit{Prefer to self-describe}. The pre-survey had six Likert scale questions, collecting participants' prior collaboration experiences and exposure to gender bias and inequality. For example, \textit{Have you experienced gender bias in the School of Computer Science}, and \textit{I am mindful of my communication approach in the classroom}? Four frequency scale questions asked participants about their role models and the amount of gender bias and inequality they previously experienced. For example, \textit{Are role models important for your discipline choices}, and \textit{Do you believe the ``boy's club'' mentality is present in the classroom}. The remaining three pre-survey questions were open-text, allowing participants to explain their prior collaboration experiences, for example, \textit{What is your level of satisfaction with the group in your last project?} 

The post-survey provided similar questions to the pre-survey but re-framed within the teamwork conducted in the study. The post-survey focused on the participants' experiences during the study, collecting their perceptions on the collaboration process and any gender bias and inequity experiences. The post-survey contained nine questions, five fewer than the pre-survey, because it did not ask for the personal details, such as name and gender. The post-survey had eight Likert scale questions to collect their teamwork experiences. For example, \textit{Did you experience gender bias while working on the project?} The one open-text question asked participants about their level of satisfaction collaborating on the software project.

The pre-post surveys were administered over Google Forms. The pre-survey was distributed the first week the teams were formed, while the post-survey was administered in the last week of the project. Participants were given a week to complete the surveys. Afterwards, the survey responses were exported from Google Forms for analysis.

\subsection{Quantitative Data Analysis} \label{data_quant_analysis}

Three quantitative analysis approaches were used in this study: median analysis, quantitative content analysis (QCA), and comparative analysis. Section \ref{survey_quant_analysis} describes how median analysis was applied to the pre-post survey responses, to analyse participants' collaboration experiences. QCA was applied to the teams' Slack messages and their Pull Requests. The QCA approach is presented in Section \ref{survey_qca_analysis}, while Section \ref{comparative_analysis_subsubsection} presents the comparative analysis used to understand the learning gains of women and men participants.

\subsubsection{Pre-Post Survey Quantitative Analysis} \label{survey_quant_analysis}

Median analysis was applied to the pre-post survey responses to analyse participants’ collaboration experiences and to compare responses by gender. The surveys were designed to determine how the participants perceived team collaboration and if their perceptions differed by gender. The Mann-Whitney U Test \citep{greasley:2008} was used to identify any statistical differences within the pre-post surveys. IBM SPSS Statistics v25 was used to perform the statistical analysis reported in this paper. The Mann-Whitney U Test enabled us to compare the participants’ changed perceptions after collaborating on the open-source project. In the pre-survey, the participants provided the number of projects they previously worked on. We used a spreadsheet to calculate the median for the prior projects. The participants’ responses were supplied as an open-text responses, so the data was first numerically quantified, for example, \textit{``six''} to \textit{6}. Median analysis was also performed on the Likert scale questions using a spreadsheet.

\subsubsection{Quantitative Content Analysis} \label{survey_qca_analysis}

Quantitative content analysis (QCA), an approach for analysing textual language in articles and transcripts \citep{riffe:2019}, was applied to Slack messages for an overview of participants' communications. To perform QCA, the communication history from the teams' Slack channels were extracted using the Slack API\footnote{https://api.slack.com/apis}. The history was saved as text files organised by teams. QCA was used to generate Slack message frequencies, presenting message frequencies by team and gender. Two Python tools were constructed to perform QCA on the teams' Slack messages. The first tool was a pre-processing tool that converted the multi-line messages to one line, allowing us to quantify the participants' Slack messages. The single lines contained the participants' ID, the date the message was sent, and the message content. The second tool aggregated the participants' messages, to import into a spreadsheet to quantify the results. Within the spreadsheet, message frequencies were quantified by the individual team and by gender across the teams. To validate the two tools, a small dataset was manually constructed by the authors using the first 50 Slack messages generated by Team-1. The dataset was used during the implementation and testing of the QCA tools, to confirm the tools were working as expected.

\begin{figure}[htp]
\centering
\begin{tabular}{p{2.5cm}p{9cm}}
\textbf{Student 4.2:}& \texttt{https://github.com/JabRef/jabref/pull/PR-4.1}\\
&Fix issue \#PR-Issue-1\\
&\makebox[0pt][l]{$\square$}\raisebox{.15ex}{\hspace{0.1em}} \hspace*{3mm} Change in CHANGELOG.md described in a way that is understandable for the average user (if applicable)\\
&\makebox[0pt][l]{$\square$}\raisebox{.15ex}{\hspace{0.1em}} \hspace*{3mm}Tests created for changes (if applicable)\\
&\makebox[0pt][l]{$\square$}\raisebox{.15ex}{\hspace{0.1em}$\checkmark$} Manually tested changed features in running JabRef (always required)\\
&\makebox[0pt][l]{$\square$}\raisebox{.15ex}{\hspace{0.1em}$\checkmark$} Screenshots added in PR description (for UI changes)\\
&\makebox[0pt][l]{$\square$}\raisebox{.15ex}{\hspace{0.1em}} \hspace*{3mm} Checked documentation: Is the information available and up to date?...\\
\end{tabular}
\caption{Example of a Pull Request Submission from Team-4. The Pull Request format is using checkboxes to denote completed and validated tasks. This format style is recommended by the JabRef project maintainers.}
\label{fig:pr_example}
\end{figure}

Within the Slack messages, we examined the teams' communication for messages related to Pull Requests (PRs). The communication might provide sight into how the teams reacted to the maintainers' feedback on their PRs. We also wanted to determine whether communication frequency could be an indicator for PR acceptance. To identify Slack messages related to the PRs, we searched the teams' Slack channels for the keywords \textit{``Pull Requests''} and \textit{``PR''}. We also searched for the PR URLs, the discussion webpage hosted on the GitHub platform. In addition to calculating the PR message frequency on the teams' Slack channels, we also calculated the message frequency on the PR discussion webpage. On this webpage, the \textit{Conversation} section contains the numeric value of messages exchanged.

To report on the PRs, we collected the message frequencies in a spreadsheet, along with the state of the PR at the end of the project: \textit{Accept}, \textit{Abandon}, or \textit{Revise}. The state is noted on the PR discussion webpage displayed at the top of the \textit{Conversation} section. The spreadsheet also collected the team member's ID that submitted the PR, their gender, and submission time. For example, the PR PR-4.1 for Team-4 was submitted by Student 4.2. Figure \ref{fig:pr_example} shows Student 4.2 letting his team know their PR was submitted to GitHub. Their initial PR message contains the issue number being addressed by the PR and an explanation on how the team fixed the issue \#PR-Issue-1. Because of the study's ethics approval, we were unable to analyse the GitHub maintainers' messages; therefore, we only report on their message frequency.
 
\subsubsection{Comparative Analysis}\label{comparative_analysis_subsubsection}

This section describes the analysis process used to evaluate the participants' learning gains, generating findings to answer the study's third research question (RQ3). We also explain our motivations for selecting comparative analysis \citep{pickvance:2001}. Comparative analysis enables us to correlate the participants' learning gains with the online communications the participants performed during the study. We measure learning gains through students' marks on course assessments. We want to determine whether there is a linear relationship between the marks awarded to the participants on course assessments and communication since communication skills contribute to students performing better in groups. Communication skills create ``a positive link between active and collaborative approaches and reported gains in problem-solving'' \citep[p.~129]{terenzini:2001}. Demonstrating this linear relationship could explain the influence communication has on participants' understanding of learning concepts and encourage knowledge transfer to other individual assessments offered in the course, such as quizzes and essays. A previous study by \citet{pepe:2012} identified a relationship between Grade Point Average (GPA) and study skills involving active participation with written communication \citep{thomas:1993}. We leverage these findings since they relate to our work using written communication during group work. However, there are different factors between this study and ours. Pepe's work was conducted in a non-computing course at a Turkish institution. The assignments and culture differ from the context of our study with Software Engineering students at an Australian institution. 

For the comparative analysis, we examine the participants' overall grades to draw any correlation between the grades and communication generated by the groups. We collected the participants' course marks from the Canvas Learning Management System (LMS), the platform used in the course to administer instructional materials and assessments. We export the marks from the LMS as a spreadsheet for analysis, which includes the marks for each assessment (See Table \ref{table_course_assessment}). We perform median analysis twofold on the collected data. The first median analysis collated the participants' marks by teams, while the second collated the marks by gender. To help illustrate the spread of participants' marks within the median by teams and gender, we calculate the standard deviation (SD) and present it alongside the median values. For the gender view, we use Cohen's \textit{d\textsubscript{s}} \citep{cohen:1988} with pooled variance to address the different group sizes. For median analysis, we remove the marks for \textit{Team communication} (20\%) from the spreadsheet to reduce bias in the results. \textit{Team communication} is an assessment marking participants’ contributions to their teams' Slack communications. Those participants receiving higher marks in this assessment demonstrate higher engagement during the group work. R software environment \citep{rfoundation:2018} is used to perform this analysis.  

The next step in the comparative analysis process evaluates two individual course assessments: a \textit{Quiz} and a \textit{Reflection essay}. These marks on these individual assessments provide insight into how team communication may influence the student's learning. Unfortunately, we cannot report the learning gains at the individual level due to the study's ethics approval. Reporting on individuals' marks could inadvertently identify them. Instead, we perform the analysis at the group level, collating the marks of team members' assessments for comparison. We used the course assessments to evaluate the impact of communication on teams and genders. The students completed these assessments at the end of the semester. The \textit{Quiz} reinforces learning concepts covered during the semester, while the \textit{Reflection essay} encourages self-reflection on contributing to the open-source project. We used Wilcoxon signed-rank test (\textit{W}) to compare with these individual assessments since the marks awarded have a non-normal distribution. We also used Cohen's \textit{d\textsubscript{s}} on the average (mean) value of the \textit{Quiz} and \textit{Reflection essay} when grouped by gender for effect size. The marks for these assessments were evaluated in a separate spreadsheet to evaluate the assessments by teams and gender. Once we completed the comparative analysis, we exported the results to report the findings. 

\subsection{Qualitative Data Analysis} \label{data_qual_analysis}

We performed a qualitative analysis of the collected data. Section \ref{survey_qual_analysis} describes the qualitative analysis conducted on the survey responses. To analyse the teams' communications, we evaluate how the students initiated and completed online team communications. We discuss how we analyse the teams' communications in two sections. The first section, Section \ref{survey_conversation_analysis_opening_sequences}, describes the analysis for the initiated communications conducted in the teams' Slack channels. Section \ref{survey_conversation_analysis_sequence_completions} describes how we analysed students' completion of these communications. Analysing the initiated (\textit{opening sequences}) and the completions (\textit{sequence completions}) generated two separate datasets. Because we coded the opening sequences and sequence completions separately and conducted the statistical analysis of the two datasets separately, we did not need to perform multiple testing correction on our results. 

\subsubsection{Pre-Post Survey Qualitative Analysis} \label{survey_qual_analysis}

Thematic content analysis \citep{marshall:1999} was performed on the open-text pre-post survey responses, using NVivo to code the responses. The questions related to participants' teamwork satisfaction and to their prior teamwork experiences. To code students' satisfaction working in teams, new nodes were created for the emerging satisfaction levels. For the question related to group formation, four new nodes were created: \textit{by the students}, \textit{by the lecturer}, \textit{a combination of both strategies}, and \textit{not applicable}. \textit{Not applicable} represents responses that do not relate to the pre-survey question. The coding of these open-text questions formed matrices extracted from NVivo and imported into SPSS for statistical analysis.

\subsubsection{Conversation Analysis - Opening Sequences} \label{survey_conversation_analysis_opening_sequences}

In this study, we used conversation analysis (CA) \citep{sacks:1984} to evaluate the participants' messages on their teams' Slack channels. The study's ethics approval limited the data analysis and collection to the students' communication in teams' Slack channels. CA has been previously used to examine the context of online asynchronous communication \citep{meredith:2019}. We used the turn-taking structure described in Section \ref{section_collab_learning} to understand team members' motivations when initiating a conversation and the methods they used to complete the conversations. CA analyses communication through \textit{turn-taking}, where a turn is a participant's action when contributing to a conversation. Successive turns create a \textit{sequence organisation}, an ordered series of events for a common action \citep{schegloff:2007}. Sometimes, sequence organisations contain \textit{disruptions}, which are turns associated with another sequence. Disruptions are common in online asynchronous communication because of the overlap during message construction \citep{garcia:1998}. 

\begin{table}[h!]
\caption{Coding Framework for Conversation Analysis} \label{coding_table} 
\begin{threeparttable}
\resizebox{\textwidth}{!}{\begin{tabular}{cp{4cm}|p{13cm}}
\multirow{12}{*}{\rotatebox[origin=c]{90}{Opening Sequence Codes}}\\
&\textbf{Behaviour} & \textbf{Definition}   \\ \cline{2-3}
&\multicolumn{2}{c}{\cellcolor[gray]{0.8}\textbf{\shortstack[l]{Opening Sequence Codes Based on Vivian et al. (2013, p. 108)}}}\\
\cline{2-3}
&Backup Behaviour* & Assists team members to help complete project tasks and processes. \\ \cline{2-3}
&Communication & Exchanges information with peers and responds to their plans and goal-setting. \\ \cline{2-3}
&Coordination & Reports on the learning activities and processes. \\ \cline{2-3}
&Feedback** & Gives, seeks, and receives information from peers on their contributions to the project. \\ \cline{2-3}
&Monitoring & Monitors team's processes, progress, and activities. \\ \cline{2-3}
&Team Leadership & Restates problem, initiates team planning, identifies items that need to be addressed.\\ \cline{2-3}
&Team Orientation & Communicates socially or the communication is a function of social communication.\\
\cline{2-3}
&\multicolumn{2}{c}{\cellcolor[gray]{0.8}\textbf{Emerging Opening Sequence Codes}}\\
\cline{2-3}
&Discussing \newline Deliverables & Works towards the completion of a presentation for the final project.\\
\cline{2-3}
&Scheduling Meeting & Organises and arranges a face-to-face or online team meeting. 
\end{tabular}}
\hspace*{5mm}

\resizebox{\textwidth}{!}{\begin{tabular}{cp{4cm}|p{13cm}}
\multirow{11}{*}{\rotatebox[origin=c]{90}{\hspace*{-2.5cm}Sequence Completion Codes}}\\
&\textbf{Behaviour} & \textbf{Definition}   \\ \cline{2-3}
&\multicolumn{2}{c}{\cellcolor[gray]{0.8}\textbf{General Sequence Completion Codes Based on \citet{hoey:2017}}}\\\cline{2-3}
&Expansion - Complete & Prepares to conclude the sequence. \\ \cline{2-3}
&Expansion - Continue & Continues the sequence with the same course of action. \\ \cline{2-3}
&Sequence Initiation & Continues the conversation to another topic. \\ \cline{2-3}
&Silence & Results in no action from participants when some action can be done. \\ \cline{2-3}
&\multicolumn{2}{c}{\cellcolor[gray]{0.8}\textbf{Sequence Recompletion Codes Based on \citet{hoey:2017}}}\\ \cline{2-3}
& Action Redoings & Same participant continues the sequence with similar topic before the lapse. \\ \cline{2-3}
& Delayed Replies & Addresses the sequence after a lapsed period of time.\\ \cline{2-3}
& Post-Sequence \newline Transition & Participants consider the lapse as an opportunity to shift the sequence's topic to something else.\\ \cline{2-3}
& Turn-exiting& A method in which a participant abandons conversing in the sequence.
\end{tabular}}
\begin{tablenotes}[flushleft]
\item \large{*} \small{Subcomponents: seeking and supporting team members.}
  \item \large{**} \small{Subcomponents: giving, receiving, and seeking feedback.}
\end{tablenotes}
\end{threeparttable}
\end{table}

To analyse the initiated communication, three researchers performed the twofold conversation analysis process outlined by \citet{woodruff:2004} to interpret the teams' initiated communication. The first step identifies the sequences through their \textit{opening sequences}, such as a greeting or topic initiation that resets the communication. We classified a disruption as an opening sequence if it did not relate to an existing sequence. In the second step, we coded the sequences using NVivo, to determine students' motivations for initiating sequences. For the analysis process, we used deductive coding with an initial coding framework previously used to analyse CS students' teamwork behaviours during online team activities \citep{vivian:2013}. The framework was established using the \citet{dickinson:2009} teamwork model, with refinements by \citet{vivian:2013} to include role behaviours. Table \ref{coding_table} shows the model's seven behaviours and two codes emerging from the coding process: \textit{Discussing Deliverables} and \textit{Scheduling Meeting}. The researchers observed student teams discussing assessment due dates and confirming the dates and times of their next face-to-face meetings. The table also shows the definitions for each behaviour. Sequences were assigned a behaviour node and a node to identify the student initiating the sequence that contained their gender and de-identified ID for anonymity. The three researchers performed a reflexive process to ensure they mitigated any biases when analysing the data.

\begin{figure}[h]
\centering
\begin{tabular}{p{2.5cm}p{6cm}}
\textbf{Student 7.1:}& Hi @Student 7.4, were you able to push the changes in git already? \smiley{}\\

\textbf{Student 7.4:}& I did, did the push not go through? ...\\
\textbf{Student 7.1:} & oh i was waiting for confirmation. I'll check it thanks\\
\end{tabular}
\caption{Example of a Sequence Organisation - Team Members are using a leave-talking statement that shows satisfaction with completing the conversation with ``thanks''}
    \label{fig:meeting_example}

\end{figure}

To demonstrate our coding process, we use Figure \ref{fig:meeting_example} as our exemplar. In this example, we assigned the opening sequence to the \textit{Monitoring} and \textit{Expansion - Complete} nodes for the sequence completion. The \textit{Student 7.1} node was also assigned, representing a woman participating in Team-7, opening and closing the sequence. The inter-rater reliability metric was used to ensure consistency in the rating system for both the opening sequence and sequence completion. At the start of the coding process, the three researchers discussed and agreed upon the coding protocol using Team-1 sequences as the basis for the protocol discussion. For the opening sequences, the decision to code against the \textit{Backup Behaviour} or \textit{Feedback} node was based on completing the task or process discussed in the sequence. Discussions on completed tasks or processes were assigned \textit{Feedback}, while in-progress work was assigned \textit{Backup Behaviour}. The researchers decided to create two emerging nodes, shown in Table \ref{coding_table}, for meetings and the deliverables because these were specific themes frequently raised in the teams' communication, and we wanted to identify themes that were of strong interest to the students. The authors also decided that sequences focusing on a common goal would be assigned to the same node. For example, sequences related to arranging or attending meetings were assigned the same node, \textit{Scheduling Meeting}. 

\subsubsection{Conversation Analysis - Sequence Completions} \label{survey_conversation_analysis_sequence_completions}

Like the qualitative analysis on the opening sequences, described in Section \ref{survey_conversation_analysis_opening_sequences}, we also use conversation analysis on the teams' \textit{sequence completions} \citep{schegloff:2007}, where the speaker and hearers, or writer and recipients in computer-mediated communications (CMC), acknowledge the end of a sequence. Sometimes, leave-talking statements, such as \textit{``bye''} and \textit{``thanks''}, denote the end of a sequence completion. Figure \ref{fig:meeting_example} shows an example of a sequence. Student 7.1 asks a peer about an outstanding task. Student 7.1 uses a leaving-talking statement, \textit{``thanks''}, to show satisfaction with completing the conversation. Within conversations, \textit{sequence recompletions} can occur, where a writer will ``bring to completion a sequence of talk that was already treated as complete'' \citep[p.~47]{hoey:2017}. Sequence recompletions can sometimes emerge when there are lapses in the conversation. Lapses in online communications can occur when writers and recipients are away from the online collaboration platform and cannot respond immediately.

Firstly, three researchers had to identify the sequence completions, using the guidance of \citet{schegloff:2007}. The guidelines state that the clear indicator of completing a sequence is the beginning of another sequence, which demonstrates the speaker's recognition that the previous sequence has concluded. Three researchers coded the sequence completions using deductive coding with an initial framework containing eight nodes, shown in Table \ref{coding_table}. Table \ref{coding_table} presents the eight  sequence completions \citep{hoey:2017}. The initial coded framework is divided into two categories: \textit{General} and \textit{Sequence Recompletion}. We used a similar node assignment as coding the opening sequences for the sequence completions. We assigned a sequence completion code with a node to identify the student writing the statement and a node to record their gender. For sequence completions assigned the \textit{Silence} node, no student was assigned.

Like the coding of the opening sequences, the three researchers also performed a reflexive process on the second dataset to mitigate any biases from coding. The authors examined the difference between \textit{Expansion - Continue} and \textit{Post-Sequence Transition}. The coding was based on how the writer acknowledged the initiated communication. If the writer included the initiated communication topic, the sequence completion was coded as \textit{Expansion - Continue}. If the writer does not acknowledge the initiated communication, the sequence completion is coded with the \textit{Post-Sequence Transition} node. To validate the coding protocols for the opening sequences and sequence completions, the three researchers coded the sequences for Team-1, where they compared the coding results for inter-rater reliability \citep{cohen:1968} using Cohen's Kappa (k) to measure agreement. The coding achieved a kappa of 0.87, an acceptable agreement rate \citep{stemler:2004}. Two researchers divided the remaining teams' Slack communications to code using the agreed-upon coding protocol. The third researcher oversaw the remaining teams coding process, to ensure the two researchers adhered to the coding protocol used on Team-1. When coding for all teams was completed for the two datasets, opening sequences and sequence completions, the primary author extracted the coded frameworks from NVivo as matrices to identify the teamwork behaviours. Pearson's chi-squared test (${\chi}^2$) was applied to the matrices, comparing the behaviours of women to men initiating and completing conversations. The data from the coded frameworks further triangulated findings from the survey responses and quantitative analysis of the Slack messages.

Also, during the coding process of the sequence completions, we examined lapses in communication. We coded sequence completions as \textit{Delayed Replies} when the responses took 60 or more minutes. We decided on 60 minutes due to prior work \citep{elmezouar:2021} identifying the response rate of professional Software Engineers at an hour while collaborating online. We coded any turns that repeated more than once in the sequence as \textit{Action Redoings}.

\section{Results} \label{section_results}

We present the results from this study in three sections. Section \ref{section_participants} presents the participants and their collaboration experiences. Section \ref{section_team_collaboration} describes the results of analysing the teams' communication over Slack and with the project maintainers. Section \ref{section_grades} presents the learning gains from the women and men participating in the study.

\begin{figure}[h]
\centering
\begin{tikzpicture}
\begin{axis}[
    ybar stacked,
    legend style={
    legend columns=1,
        at={(1.00,1)},
        draw=none,
        area legend,
    },
    ytick=data,
    tick label style={font=\footnotesize},
    legend style={font=\footnotesize},
    label style={font=\footnotesize},
    ytick={0,100,300,500,700,900,1100,1300,1500},
    height=170,
    width=.9\textwidth,
    bar width=4mm,
    ylabel={Slack Messages (\#)},
    x tick label style={rotate=45, anchor=north east, inner sep=0mm},
    xtick=data,
    xticklabels={Team-1, Team-2, Team-3, Team-4\textsuperscript{*}, Team-5, Team-6, Team-7, Team-8\textsuperscript{*}},
    ymin=0,
    area legend,
]
\addplot[female,fill=female] coordinates
{(1,0) (2,54) (3,0) (4,0) (5,1454) (6,8) (7,453) (8,117)};
\addplot[male,fill=male] coordinates
{(1,531) (2,287) (3,58) (4,375) (5,0) (6,72) (7,199) (8,273)};
\legend{Women, Men}
\end{axis}  
\end{tikzpicture}

\setlength\tabcolsep{4pt}
\begin{tabular}{|c|c|c|c|c|c|c|c|}
\hline
\multicolumn{2}{|c|}{\cellcolor[gray]{0.8}\textbf{\small{Team-1}}} & \multicolumn{2}{c|}{\cellcolor[gray]{0.8}\textbf{\small{Team-2}}} & \multicolumn{2}{c|}{\cellcolor[gray]{0.8}\textbf{\small{Team-3}}}& \multicolumn{2}{c|}{\cellcolor[gray]{0.8}\textbf{\small{Team-4*}}} \\
\multicolumn{2}{|c|}{\cellcolor[gray]{0.8}\textbf{\small{Zettlr}}} & \multicolumn{2}{c|}{\cellcolor[gray]{0.8}\textbf{\small{Pygments}}} & \multicolumn{2}{c|}{\cellcolor[gray]{0.8}\textbf{\small{Pygments}}}& \multicolumn{2}{c|}{\cellcolor[gray]{0.8}\textbf{\small{JabRef}}} \\\hline
\textbf{\small{Women}} & \textbf{\small{Men}} &  \textbf{\small{Women}} & \textbf{\small{Men}} & \textbf{\small{Women}} & \textbf{\small{Men}} & \textbf{\small{Women}} & \textbf{\small{Men}} \\ \hline
0&5&1&4&0&6&0&4\\ \hline
\multicolumn{2}{|l|}{UG=4 \hspace*{4mm}PG=1} &
\multicolumn{2}{|l|}{UG=1 \hspace*{4mm}PG=4} & 
\multicolumn{2}{|l|}{UG=6 \hspace*{4mm}PG=0} & 
\multicolumn{2}{|l|}{UG=1 \hspace*{4mm}PG=3} \\

\hline
\multicolumn{8}{l}{} \\
\hline
\multicolumn{2}{|c|}{\cellcolor[gray]{0.8}\textbf{\small{Team-5}}} & \multicolumn{2}{c|}{\cellcolor[gray]{0.8}\textbf{\small{Team-6}}} & \multicolumn{2}{c|}{\cellcolor[gray]{0.8}\textbf{\small{Team-7}}}& \multicolumn{2}{c|}{\textbf{\cellcolor[gray]{0.8}\small{Team-8*}}} \\ 
\multicolumn{2}{|c|}{\cellcolor[gray]{0.8}\textbf{\small{Zettlr}}} & \multicolumn{2}{c|}{\cellcolor[gray]{0.8}\textbf{\small{Zettlr}}} & \multicolumn{2}{c|}{\cellcolor[gray]{0.8}\textbf{\small{JabRef}}}& \multicolumn{2}{c|}{\cellcolor[gray]{0.8}\textbf{\small{Pygments}}} \\
\hline
\textbf{\small{Women}} & \textbf{\small{Men}} & \textbf{\small{Women}} & \textbf{\small{Men}} & \textbf{\small{Women}} & \textbf{\small{Men}} & \textbf{\small{Women}} & \textbf{\small{Men}}  \\ \hline
4&0&1&5&2&3&1&3 \\ \hline
\multicolumn{2}{|l|}{UG=1 \hspace*{4mm}PG=3} & 
\multicolumn{2}{|l|}{UG=0 \hspace*{4mm}PG=6} &
\multicolumn{2}{|l|}{UG=1 \hspace*{4mm}PG=4} &
\multicolumn{2}{|l|}{UG=4 \hspace*{4mm}PG=0} \\
\hline
\end{tabular}
    \caption{Team Composition and Message Frequency}
    \label{fig:teamcompositions}
\captionsetup{justification=centering}
\begin{center}
* \small{Has a non-participant team member; excluded from study}
\end{center}
\end{figure}

\subsection{Participants} \label{section_participants}

Eight (53.33\%) out of 15 teams from the course volunteered for the study. There were 39 (49.37\%) student participants: nine women (23.08\%) and 30 men (76.92\%). None of the participants identified as non-binary in the survey. Of the nine women, two (22.22\%) are undergraduates, and seven (77.78\%) are postgraduates. Of the 30 men, 16 (53.33\%) are undergraduates, and 14 (46.67\%) are postgraduates. The participation of women in Software Engineering studies evaluating gender (20\% \citep{capretz:2003}, 18.6\% \citep{russo:2022}) have comparable participation rates by the genders to our research. The women had low representation in our study due to low enrolment (18.00\%) in the course. However, most (64.29\%) of the enrolled women participated in the study. The Mann-Whitney U Tests (\textit{p}$<$0.05 two-tailed) showed no significant differences between the responses to the pre-post surveys by the men and women participating in the study. The overall effect sizes for the pre-post surveys are small, where the pre-survey had an eta squared ($\eta$\textsuperscript{2}) value of 0.001, while the post-survey has an $\eta$\textsuperscript{2} value of 0.014. 

Figure \ref{fig:teamcompositions} shows the gender composition of the eight teams, containing one all-woman, three all-man, and four mixed-gender teams. Teams 4 and 8 each had a team member who did not volunteer for the study, so these team members' Slack communication was excluded from the study. The figure shows the number of under- (UG) and postgraduate (PG) students on each team and the GitHub projects they worked on. Two teams (25.00\%, teams 4 and 7) contributed to JabRef, while three (37.50\%, teams 2, 3, and 8) worked on Pygments, and the remaining three (37.50\%, teams 1, 5, and 6) teams contributed to the Zettlr project.

The majority (75\%, n=6) of the teams were formed by the students, except for Team-1 (5 men) and Team-4 (4 men), where the lecturer randomly assigned students to increase the team size. At the university, the educators typically assist students when they cannot find a team to join. In this case, the educator helped students join Team-2 and Team-4 since they had the smallest number of members. When asked about prior team formation, 18 (46.15\%) participants stated that the students formed all their previous teams, 15 (38.46\%) experienced group formation by students and lecturers, and five (12.82\%) participants had their teams formed by educators. We could not identify one (2.57\%) participant's response that answered with \textit{``by discussion''}.

From the survey results, the participants had prior teamwork experience (median 5 team projects). The women had more ($\mu$=8 projects) group experience than the men participating in the study ($\mu$=6 projects). All participants had prior experience collaborating in teams. Most (92.31\%, n=36) of the participants recalled positive experiences in their past projects, except three (7.70\%, W=1, M=1) participants expressed dissatisfaction. Participants voicing dissatisfaction did not mention bias as a reason for their sentiments. For example, one woman expressed dissatisfaction, stating, \textit{``My last project involved 8 team members (software engineering project, which is a third year subject). It was overwhelming and I'd say I was dissatisfied with my team''}. One man expressing satisfaction with prior collaborative projects noted a gender imbalance in the teams, stating, \textit{``Satisfied, although there was a gender imbalance (no girls) and one of the members dropped without saying anything''}; however, we cannot discern the reasons for the participant providing this additional information, and it does not raise biases in the team for the individual's motivation for leaving the course.

When asked about gender bias experience, seven (17.95\%, Women=2, Men=5) participants had previously experienced gender bias, 21 (53.85\%, Women=6, Men=15) had not, and 11 (28.20\%, Women=1, Men=10) were unsure. For the teamwork in this study, the majority (66.67\%, n=26) of the participants were satisfied with working in their teams. The most common response (n=18, 46.15\%) was \textit{``very satisfied''}. One woman acknowledged difficulties collaborating at times in the team, stating, \textit{``I'd rate it 8/10 as we did end up finishing what we intended to do but there were bumpy days''}. However, we cannot discern whether her comment on \textit{``bumpy days''} are due to biases experienced in the teams. The most negative response from the participants was a man responding \textit{``neutral''} to the question.

As for working on the project during the study, the participants had positive experiments. For example, when asked how they conducted themselves during the group projects, women were slightly more mindful ($\mu$=4) in their communication approach than the men ($\mu$=3.57). Still, their responses were not statistically significant (\textit{p}=0.0703 two-tailed) in claiming differences in the genders. Almost all (97.44\%, n=38) participants felt included in their group projects. Overall, we found the teams responding to the pre-post survey questions similarly with no outliers to warrant further investigation.  

\subsection{Team Collaboration} \label{section_team_collaboration}

\begin{figure}[h!]
\centering
\includegraphics[width=1\textwidth]{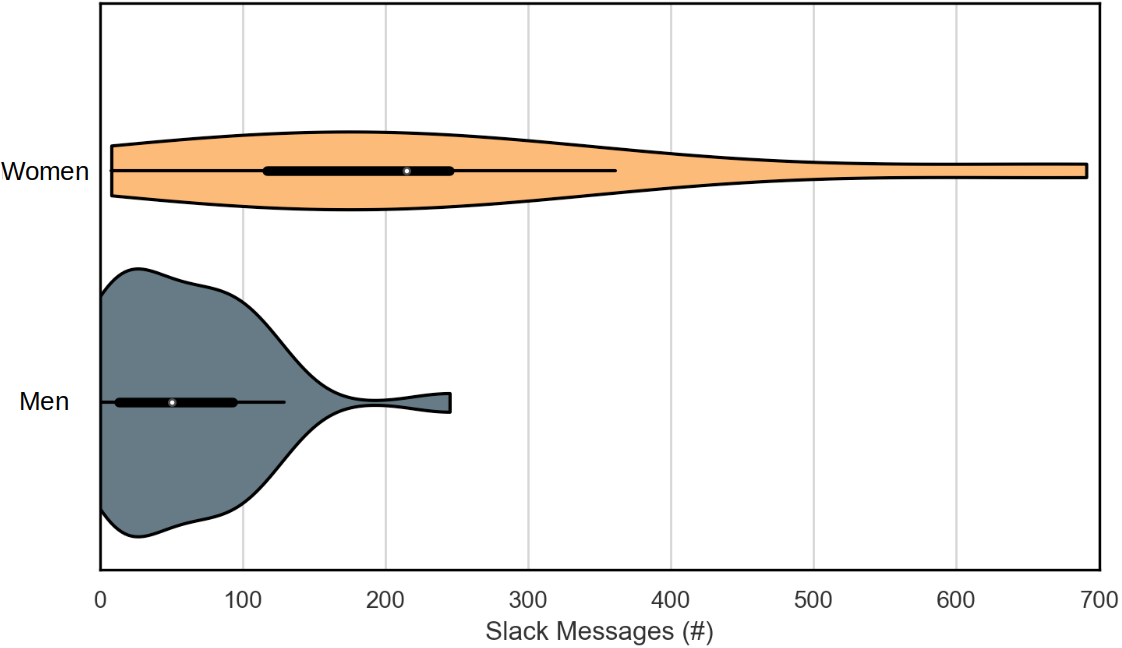}
\caption{Slack Message Frequency by Gender}
\label{fig:genderMessages}
\end{figure}

This section presents the results of the analysis of the teams' Slack communication. In total, the eight teams generated 3881 Slack messages. Figure \ref{fig:teamcompositions} shows the message frequency for each team, grouping the messages by gender, where the all-woman team, Team-5, generated the most (37.46\%, ${f}$=1454) messages. When examining the women's communication within mixed-gender teams, the results show higher communication frequency within the team with more than one woman. Team-7 (W=2, M=3) showed the women in the team initiated 69.48\% of the communication, but in Team-6 (W=1, M=5), one woman on the team initiated 7.50\% of the communication. We cannot discern whether the undergraduate (UG) and postgraduate (PG) studies influenced teams' communications. For example, we observe the two teams with the lowest message frequencies are comprised of students at different graduate levels. Team-3 (UG=6, PG=0, ${f}$=58 messages) comprises all undergraduate students, while Team-6 (UG=0, PG=6, ${f}$=80 messages) comprises all graduate-level students. More research is necessary to investigate the impact students' graduate level has on their communication during collaborative learning.

Figure \ref{fig:genderMessages} shows another view of the message frequency, demonstrating women generating more messages. The nine women generated more messages (2086 messages, $\bar{x}$=231.78) than the 30 men (1795 messages, $\bar{x}$=64.11). This figure shows all the women communicating with their teams, while two men did not contribute to their teams' communication over Slack.

For the remainder of this section, we divide the presentation of the results into three areas. Section \ref{subsubsection_teams_initiated_communication} presents the teams' initiated communication behaviours, while Section \ref{subsubsection_teams_communication_reactions} presents how the teams reacted to the initiated communications. Section \ref{subsub_pull_requests} shows the results of the teams' communication about the Pull Requests.

\begin{sidewaystable}
\begin{threeparttable}
\caption{Coding Framework - Initiated Teamwork Behaviours Organised by Frequency (\showblankfemalewhitebar{0.5} Women \showblankmalewhitebar{0.5} Men. *Statistically significant differences (${\chi}^2$ $\geq$ 5 at \textit{p} $\leq$ 0.05) in initiated behaviours between women and men)}
\label{tbl:collab_results}
\begin{tabular}{p{3.5cm}|p{10mm}|p{13mm}||p{15mm}||p{10mm}|p{10mm}|p{10mm}||p{14mm}|p{14mm}|p{14mm}|p{12mm}}
\multicolumn{3}{c||}{}&\textbf{\tiny{All-Women}}&\multicolumn{3}{c||}{\textbf{\tiny{All-Man}}}&\multicolumn{4}{c}{\textbf{\tiny{Mixed-Gender}}}\\
\thead{\textbf{Behaviour} \\ (${\chi}^2$, \textit{p}-value)}&\thead{\textbf{Total} \\ (\%)} &\thead{\textbf{Total} \\ (${f}$)} &
\thead{\textbf{\scriptsize{Team-5}}\\ \scriptsize{W=4}} & 
\vspace*{-9mm}\thead{\textbf{\scriptsize{Team-4}}\\ \scriptsize{M=4}} &\vspace*{-9mm}\thead{\textbf{\scriptsize{Team-1}} \\ \scriptsize{M=5}} & \vspace*{-9mm}\thead{\textbf{\scriptsize{Team-3}}\\ \scriptsize{M=6}} & 
\vspace*{-9mm}\thead{\textbf{\scriptsize{Team-7}}\\ \scriptsize{W=2, M=3}} & \vspace*{-9mm}\thead{\textbf{\scriptsize{Team-8}}\\ \scriptsize{W=1, M=3}}& \vspace*{-9mm}\thead{\textbf{\scriptsize{Team-2}}\\ \scriptsize{W=1, M=4}} & 
\vspace*{-9mm}\thead{\textbf{\scriptsize{Team-6}}\\ \scriptsize{W=1, M=5}}
\\ \hline

\hspace*{-2mm}\small{\textbf{Communication}\newline\newline\hspace*{-2mm}${\chi}^2$=1.688, \textit{p}=0.4230}& \small{33.16\%} &\small{${f}$=124} \newline \femalewhitebar{0.43} \newline \malewhitebar{0.57} & \small{${f}$=32} \newline \femalewhitebar{1.0} & \small{${f}$=15} \newline \newline\malewhitebar{1.0}& \small{${f}$=12} \newline \newline\malewhitebar{1.0} & \small{${f}$=5} \newline \newline\malewhitebar{1.0}& \small{${f}$=28} \newline \femalewhitebar{0.64}\newline \malewhitebar{0.36}& \small{${f}$=13} \newline \femalewhitebar{0.15}\newline \malewhitebar{0.85}& \small{${f}$=17} \newline \femalewhitebar{0.06}\newline \malewhitebar{0.94}& \small{${f}$=2 } \newline\femalewhitebar{0.0}\newline\malewhitebar{1.0} \\ \hline

\hspace*{-2mm}\small{\textbf{Scheduling\newline \hspace*{-2mm}Meeting} \newline\hspace*{-2mm}${\chi}^2$=0.501, \textit{p}=0.7784}& \small{16.58\%}& \small{${f}$=62} \newline\femalewhitebar{0.48} \newline\malewhitebar{0.52}& \small{${f}$=17} \newline \femalewhitebar{1.0}& \small{${f}$=11} \newline \newline\malewhitebar{1.0}& \small{${f}$=6} \newline\newline \malewhitebar{1.0}& \small{${f}$=2} \newline\newline \malewhitebar{1.0}& \small{${f}$=14} \newline \femalewhitebar{0.71}\newline \malewhitebar{0.29}& \small{${f}$=3} \newline \femalewhitebar{0.67} \newline \malewhitebar{0.33}& \small{${f}$=6} \newline \femalewhitebar{0.17} \newline \malewhitebar{0.83}& \small{${f}$=3} \newline\femalewhitebar{0.0}\newline\malewhitebar{1.0} \\ \hline

\hspace*{-2mm}\textbf{\small{Team Leadership}}\large{*}\newline \small{\newline\hspace*{-2mm}${\chi}^2$=14.199, \textit{p}=0.0008}& \small{12.57\%} & \small{${f}$=47}  \newline\femalewhitebar{0.83} \newline \malewhitebar{0.17}& \small{${f}$=15} \newline \femalewhitebar{1.0} & \small{${f}$=1} \newline \newline\malewhitebar{1.0}& \small{${f}$=1} \newline \newline\malewhitebar{1.0}& \small{${f}$=1}\newline \newline \malewhitebar{1.0}& \small{${f}$=21} \newline \femalewhitebar{1.0}\newline \malewhitebar{0.0}& \small{${f}$=3} \newline \femalewhitebar{1.0}\newline \malewhitebar{0.0}& \small{${f}$=5} \newline\femalewhitebar{0.0}\newline\malewhitebar{1.0}& \small{${f}$=0}  \\ \hline

\hspace*{-2mm}\textbf{\small{Coordination}\large{*}} \small{\newline\newline\hspace*{-2mm}${\chi}^2$=8.598, \textit{p}=0.0136}& \small{12.03\%} & \small{${f}$=45} \newline\femalewhitebar{0.53}\newline\malewhitebar{0.47}& \small{${f}$=12} \newline \femalewhitebar{1.0} & \small{${f}$=5} \newline \newline\malewhitebar{1.0}& \small{${f}$=2} \newline\newline \malewhitebar{1.0}& \small{${f}$=2} \newline\newline \malewhitebar{1.0}& \small{${f}$=10} \newline \femalewhitebar{0.80}\newline \malewhitebar{0.20}& \small{${f}$=10} \newline \femalewhitebar{0.30} \newline \malewhitebar{0.70} & \small{${f}$=3} \newline \femalewhitebar{0.0}\newline \malewhitebar{1.0}& \small{${f}$=1} \newline \femalewhitebar{1.0}\newline \malewhitebar{0.0}\\ \hline

\hspace*{-2mm}\small{\textbf{Discussing \newline \hspace*{-2mm}Deliverables} \newline\hspace*{-2mm}${\chi}^2$=1.981, \textit{p}=0.3714} & \small{10.16\%} & \small{${f}$=38} \newline\femalewhitebar{0.47}\newline\malewhitebar{0.53}& \small{${f}$=11} \newline \femalewhitebar{1.0}  & \small{${f}$=5} \newline \newline\malewhitebar{1.0}& \small{${f}$=3} \newline\newline \malewhitebar{1.0}& \small{${f}$=3} \newline\newline \malewhitebar{1.0}& \small{${f}$=6} \newline \femalewhitebar{0.83}\newline \malewhitebar{0.17}& \small{${f}$=6} \newline \femalewhitebar{0.33}\newline \malewhitebar{0.67}& \small{${f}$=4} \newline \femalewhitebar{0.0}\newline \malewhitebar{1.0}& \small{${f}$=0} \\ \hline

\hspace*{-2mm}\textbf{\small{Monitoring}\large{*}} \small{\newline\newline\hspace*{-3mm}${\chi}^2$=21.277, \textit{p}=0.00002} &\small{6.15\%}& \small{${f}$=23}  \newline\femalewhitebar{0.70}\newline\malewhitebar{0.30}& \small{${f}$=9} \newline \femalewhitebar{1.0} & \small{${f}$=1} \newline \newline\malewhitebar{1.0}& \small{${f}$=5} \newline \newline\malewhitebar{1.0}& \small{${f}$=0} & \small{${f}$=6} \newline \femalewhitebar{1.0}\newline \malewhitebar{0.0}& \small{${f}$=2} \newline \femalewhitebar{0.50}\newline \malewhitebar{0.50}& \small{${f}$=0} \newline & \small{${f}$=0}  \\ \hline

\hspace*{-2mm}\textbf{\small{Backup Behaviour: \newline \hspace*{-2mm}Seeking}\large{*}} \small{\newline\hspace*{-2mm}${\chi}^2$=8.547, \textit{p}=0.0139} & \small{5.61\%} & \small{${f}$=21 } \newline\femalewhitebar{0.71}\newline\malewhitebar{0.29}& \small{${f}$=13} \newline \femalewhitebar{1.0}& \small{${f}$=2} \newline\newline \malewhitebar{1.0}& \small{${f}$=1} \newline\newline \malewhitebar{1.0}& \small{${f}$=0} & \small{${f}$=3} \newline \femalewhitebar{0.33}\newline \malewhitebar{0.67}& \small{${f}$=1} \newline \femalewhitebar{0.0}\newline \malewhitebar{1.0}& \small{${f}$=1} \newline \femalewhitebar{1.0}\newline \malewhitebar{0.0}& \small{${f}$=0} \\ \hline

\hspace*{-2mm}\textbf{\small{Team Orientation}\newline\newline\hspace*{-2mm}}\small{${\chi}^2$=3.009, \textit{p}=0.2221} & \small{3.74\%} & \small{${f}$=14} \newline\femalewhitebar{0.64}\newline\malewhitebar{0.36}& \small{${f}$=9} \newline \femalewhitebar{1.0} & \small{${f}$=1} \newline \newline\malewhitebar{1.0}& \small{${f}$=0} & \small{${f}$=0} & \small{${f}$=0} & \small{${f}$=2} \newline \femalewhitebar{0.0}\newline \malewhitebar{1.0} & \small{${f}$=2} \newline \femalewhitebar{0.0}\newline \malewhitebar{1.0}& \small{${f}$=0} \\ \hline\hline

\multicolumn{2}{p{4cm}|}{\hspace*{-2mm}\small{\textbf{Total Behaviours} \newline \textbf{\hspace*{-2mm}by Teams} (\%)}} &\small{${f}$=374} 
\newline\newline\femalewhitebar{0.54}\newline\malewhitebar{0.46}& \small{${f}$=118 \newline 31.55\%} \newline \femalewhitebar{1.0} & \small{${f}$=41 \newline 10.96\%} \newline\newline \malewhitebar{1.0}& \small{${f}$=30 \newline 8.02\%} \newline\newline \malewhitebar{1.0}& \small{${f}$=13 \newline 3.48\%} \newline\newline \malewhitebar{1.0}& \small{${f}$=88 \newline 23.53\%} \newline \femalewhitebar{0.78}\newline \malewhitebar{0.22}& \small{${f}$=40 \newline 10.70\%} \newline \femalewhitebar{0.28}\newline \malewhitebar{0.72}& \small{${f}$=38 \newline 10.16\%} \newline \femalewhitebar{0.08}\newline \malewhitebar{0.92}& \small{${f}$=6 \newline 1.60\%} \newline \femalewhitebar{0.17}\newline \malewhitebar{0.83}\\

\end{tabular}
\end{threeparttable}
\end{sidewaystable}

\subsubsection{Teams' Initiated Communication} \label{subsubsection_teams_initiated_communication}

The results from the conversation analysis identified eight teamwork behaviours for the opening sequences, shown in Table \ref{tbl:collab_results}. The table arranges the behaviours by frequency and shows the teams' application of the behaviours. The bar graphs illustrate the gender initiating the behaviours. The bottom row in the table represents the total behaviours generated by each team. Six of the behaviours come from the initial coding framework, while the remaining two are emerging: \textit{Scheduling Meeting} (16.58\%) and \textit{Discussing Deliverables} (10.16\%). The behaviours within \textit{Scheduling Meeting} closely relate to \textit{Team Leadership}, while those in \textit{Discussing Deliverables} have attributes of \textit{Communication}, \textit{Coordination}, and \textit{Team Leadership} behaviours. 

The most frequent behaviour was \textit{Communication} (33.16\%), for example, \textit{``just noting down our preferred issues.''} \textit{Team Orientation} (3.74\%), for example, \textit{``Girls, my laptop died''}, was rarely applied by the teams. Though unrelated to project development, \textit{Team Orientation} provides team cohesion through social communication \citep{dickinson:2009}. The sparse application of \textit{Team Orientation} is supported by previous findings \citep{adel:2011} that show students rarely engage in social interactions during online teamwork.

The Pearson's chi-squared tests on the initiated communication dataset showed statistical differences, where ${\chi}^2$ $\geq$ 7 is significant at \textit{p} $\leq$ 0.05 for four behaviours: \textit{Team Leadership}, \textit{Monitoring}, \textit{Coordination}, and \textit{Backup Behaviour: Seeking}. A reason for the statistical difference is that women predominately initiate these four behaviours. \textit{Team Leadership} (12.57\%, ${f}$=47) provided direction for the team, \textit{Monitoring} (6.15\%, ${f}$=23) showed team members asking their peers for updates on their tasks, and \textit{Coordination} (12.03\%, ${f}$=45) involved team members asking their peers to perform tasks, such as \textit{``@Student 7.2, please help perform another round of QAT (quality assurance testing)''}. \textit{Backup Behaviour: Seeking} (5.61\%, ${f}$=21) showed a team member asking for help from peers on their tasks. For example, \textit{``@Student 5.3 could you help me with my error''}.

\begin{sidewaystable}
\begin{threeparttable}
\centering
\captionsetup{justification=centering}
\caption{Coding Framework - Sequence Completion for Teams' Communication Organised by Frequency (\showblankfemalewhitebar{0.5} Women \showblankmalewhitebar{0.5} Men)}
\label{tbl:sequence_completion}
\begin{tabular}{p{3.5cm}|p{11mm}|p{13mm}||p{15mm}||p{10mm}|p{10mm}|p{10mm}||p{14mm}|p{14mm}|p{14mm}|p{14mm}}
\multicolumn{3}{c||}{}&\textbf{\tiny{All-Women}}&\multicolumn{3}{c||}{\textbf{\tiny{All-Man}}}&\multicolumn{4}{c}{\textbf{\tiny{Mixed-Gender}}}\\
\thead{\textbf{Behaviour} \\ (${\chi}^2$, \textit{p}-value)}&\thead{\textbf{Total} \\ (\%)} &\thead{\textbf{Total} \\ (${f}$)} &
\thead{\textbf{\scriptsize{Team-5}}\\ \scriptsize{W=4}} & 
\vspace*{-9mm}\thead{\textbf{\scriptsize{Team-4}}\\ \scriptsize{M=4}} & \vspace*{-9mm}\thead{\textbf{\scriptsize{Team-1}}\\ \scriptsize{M=5}} & \vspace*{-9mm}\thead{\textbf{\scriptsize{Team-3}}\\ \scriptsize{M=6}} & 
\vspace*{-9mm}\thead{\textbf{\scriptsize{Team-7}}\\ \scriptsize{W=2, M=3}} & \vspace*{-9mm}\thead{\textbf{\scriptsize{Team-8}}\\ \scriptsize{W=1, M=3}}& \vspace*{-9mm}\thead{\textbf{\scriptsize{Team-2}}\\ \scriptsize{W=1, M=4}} & 
\vspace*{-9mm}\thead{\textbf{\scriptsize{Team-6}}\\ \scriptsize{W=1, M=5}}
\\ \hline

\multicolumn{11}{c}{\cellcolor[gray]{0.8}\textbf{\small{Initial Sequence Completion Codes Based on \citet{hoey:2017}}}}\\\hline

\hspace*{-2mm}\small{\textbf{Expansion -\newline \hspace*{-2mm}Complete}}\newline\hspace*{-2mm}${\chi}^2$=1.688, \textit{p}=0.1938& 45.98\% &\small{${f}$=177} \newline \femalewhitebar{0.60} \newline \malewhitebar{0.40} & \small{${f}$=77} \newline \femalewhitebar{1.0} & \small{${f}$=13} \newline \newline\malewhitebar{1.0}& \small{${f}$=9} \newline \newline\malewhitebar{1.0} & \small{${f}$=4} \newline \newline\malewhitebar{1.0}& \small{${f}$=45} \newline \femalewhitebar{0.47}\newline \malewhitebar{0.53}& \small{${f}$=17} \newline \femalewhitebar{0.12}\newline \malewhitebar{0.88}& \small{${f}$=11 } \newline\femalewhitebar{0.45}\newline\malewhitebar{0.55}& \small{${f}$=1} \newline \femalewhitebar{0.0}\newline \malewhitebar{1.0} \\ \hline

\hspace*{-2mm}\small{\textbf{Expansion -\newline \hspace*{-2mm}Continue}\large{*}} \newline\hspace*{-2mm}${\chi}^2$=7.557, \textit{p}=0.0060& 15.32\%& \small{${f}$=59} \newline\femalewhitebar{0.73} \newline\malewhitebar{0.27}& \small{${f}$=32} \newline \femalewhitebar{1.0}& \small{${f}$=1} \newline \newline\malewhitebar{1.0}& \small{${f}$=5} \newline\newline \malewhitebar{1.0}& \small{${f}$=0} & \small{${f}$=13} \newline \femalewhitebar{0.62}\newline \malewhitebar{0.38}& \small{${f}$=4} \newline \femalewhitebar{0.50} \newline \malewhitebar{0.50}& \small{${f}$=4} \newline\femalewhitebar{0.75}\newline\malewhitebar{0.25}& \small{${f}$=0} \\ \hline

\hspace*{-2mm}\textbf{\small{Silence}}& 14.55\% & \small{${f}$=56}  & \small{${f}$=6} & \small{${f}$=11} & \small{${f}$=4} & \small{${f}$=6}& \small{${f}$=15} & \small{${f}$=7} & \small{${f}$=6} & \small{${f}$=1} \\ \hline

\multicolumn{11}{c}{\cellcolor[gray]{0.8}\textbf{\small{Sequence Recompletion Methods Based on \citet{hoey:2017}}}}\\\hline

\hspace*{-2mm}\textbf{\small{Post-Sequence}\newline \hspace*{-2mm}Transition} \small{\newline\hspace*{-2mm}${\chi}^2$=2.696, \textit{p}=0.1006} &11.17\%& \small{${f}$=43}  \newline\femalewhitebar{0.05}\newline\malewhitebar{0.95}& \small{${f}$=0}& \small{${f}$=7} \newline \newline\malewhitebar{1.0}& \small{${f}$=12} \newline \newline\malewhitebar{1.0}& \small{${f}$=3} \newline \newline\malewhitebar{1.0}& \small{${f}$=2} \newline \femalewhitebar{0.50}\newline \malewhitebar{0.50}& \small{${f}$=5} \newline \femalewhitebar{0.80}\newline \malewhitebar{0.20}& \small{${f}$=11}\newline \femalewhitebar{0.00}\newline \malewhitebar{1.00}&\small{${f}$=3} \newline \femalewhitebar{0.00}\newline \malewhitebar{1.00}  \\ \hline

\hspace*{-2mm}\small{\textbf{Delayed \newline \hspace*{-2mm}Replies} \newline\hspace*{-2mm}${\chi}^2$=0.286, \textit{p}=0.5927} & 6.23\% & \small{${f}$=24} \newline\femalewhitebar{0.21}\newline\malewhitebar{0.79}& \small{${f}$=4} \newline \femalewhitebar{1.0}  & \small{${f}$=4} \newline \newline\malewhitebar{1.0}& \small{${f}$=1} \newline\newline \malewhitebar{1.0}& \small{${f}$=0} & \small{${f}$=4} \newline \femalewhitebar{0.25}\newline \malewhitebar{0.75}& \small{${f}$=5} \newline \femalewhitebar{0.0}\newline \malewhitebar{1.0}& \small{${f}$=4} \newline \femalewhitebar{0.0}\newline \malewhitebar{1.0}& \small{${f}$=2} \newline \femalewhitebar{0.0}\newline \malewhitebar{1.0}\\ \hline

\hspace*{-2mm}\textbf{\small{Action \newline \hspace*{-2mm}Redoings\large{*}}} \small{\newline\hspace*{-2mm}${\chi}^2$=5.850, \textit{p}=0.0156}& 5.71\% & \small{${f}$=22} \newline\femalewhitebar{0.73}\newline\malewhitebar{0.27}& \small{${f}$=12} \newline \femalewhitebar{1.0} & \small{${f}$=2} \newline \newline\malewhitebar{1.0}& \small{${f}$=1} \newline\newline \malewhitebar{1.0}& \small{${f}$=0} & \small{${f}$=2} \newline \femalewhitebar{1.0}\newline \malewhitebar{0.0}& \small{${f}$=3} \newline \femalewhitebar{0.33} \newline \malewhitebar{0.67} & \small{${f}$=1} \newline \femalewhitebar{0.0}\newline \malewhitebar{1.0}& \small{${f}$=1} \newline \femalewhitebar{0.0}\newline \malewhitebar{1.0}\\ \hline

\hspace*{-2mm}\textbf{\small{Turn-Exiting}} \small{\newline\newline\hspace*{-2mm}${\chi}^2$=0.000, \textit{p}=1.0000} & 1.04\% & \small{${f}$=4 } \newline\femalewhitebar{0.25}\newline\malewhitebar{0.75}& \small{${f}$=0} & \small{${f}$=1} \newline\newline \malewhitebar{1.0}& \small{${f}$=0} & \small{${f}$=0} & \small{${f}$=1} \newline \femalewhitebar{0.0}\newline \malewhitebar{1.0}& \small{${f}$=2} \newline \femalewhitebar{0.50}\newline \malewhitebar{0.50}& \small{${f}$=0} & \small{${f}$=0} \\ \hline

\multicolumn{2}{p{3.5cm}|}{\hspace*{-2mm}\small{\textbf{Total Completions} \newline \textbf{\hspace*{-2mm}by Teams} (\%)}} &\small{${f}$=385} & \small{${f}$=131 \newline 34.03\%}& \small{${f}$=39 \newline 10.13\%} & \small{${f}$=32 \newline 8.31\%} & \small{${f}$=13 \newline 3.38\%} & \small{${f}$=82 \newline 21.30\%} & \small{${f}$=43 \newline 11.17\%} & \small{${f}$=37 \newline 9.61\%} & \small{${f}$=8 \newline 2.07\%} \\

\end{tabular}
\begin{tablenotes}
  \item \hspace{-5mm}\large{*} \small{Statistically significant differences (${\chi}^2$ $\geq$ 5 at \textit{p} $\leq$ 0.05) in the sequence completion behaviours between the women and men} 
\end{tablenotes}
\end{threeparttable}
\end{sidewaystable}

\begin{figure}[h]
\centering
\begin{tabular}{p{2.5cm}p{9cm}}

\textbf{Student 1.5:}& Can we meet tomorrow? Maybe 10am would be best for me?\\
& because I have another group meeting at 11, and then other commitments \\
\textbf{Student 1.1:}& Sure \\ 
\textbf{Student 1.2:}& \faThumbsOUp \\
\textbf{Student 1.3:}& \faThumbsOUp \\

\end{tabular}
\caption{Example of \textit{Expansion - Complete} - Team members using this sequence completion to confirm the conclusion of the sequence}
    \label{fig:expansion_complete_completion_example}
\end{figure}

\subsubsection{Teams' Reactions to Communication}\label{subsubsection_teams_communication_reactions}

This section presents how teams reacted to communications initiated in Slack by team members. We present the results in Table \ref{tbl:sequence_completion}, analysing the sequence completions of teams' communication and presenting them by highest frequency. The analysis produced seven behaviours from the initial coding framework. No emerging nodes were identified during the coding process. The initial coding framework contained eight behaviours collated into two categories: \textit{General} and \textit{Sequence Recompletion}. Most (n=291, 75.39\%) of the coded sequence completions are from the \textit{General} category. The results show \textit{Expansion - Complete} (45.98\%, n=177) was the most applied code, while \textit{Turn-Exiting} was infrequently applied (1.04\%, n=4). Figure \ref{fig:expansion_complete_completion_example} shows an example of \textit{Expansion - Complete}, where a participant initiates a sequence to set up a meeting, including emoji responses from team members to confirm the meeting time.

The initial coding framework contained eight behaviours, but the \textit{Sequence Initiation} node was not identified during the coding process. The \textit{Sequence Initiation} node represents a topic change in the conversation. We interpreted the change in the conversation as a new opening sequence, which explains the node's absence in the coding results. Our results show the frequency of initiated (${f}$=374) and completed (${f}$=385) communications are not aligned due to how team members responded to the initiated communications. For example, Figure \ref{fig:action_redoings_sequence_completion_example} shows how \textit{Action Redoings} generates multiple closing sequences. In this example, Student 7.3 (Woman) is monitoring the team's progress. Student 7.3 is using the Slack reserved term, \texttt{@channel}, to send an announcement to all the team members. The second \textit{Action Redoing} shows Student 7.3 addressing two team members, @Student 7.2 (Man) and @Student 7.4 (Man) to monitor their progress.

\begin{figure}[t!]
\centering
\begin{tabular}{p{2.5cm}p{9cm}}

\textbf{Student 7.3:}& Hi @channel, please update the task progress for tracking thanks!\\
\textbf{Student 7.3:}& @channel, Hi Team! just wanna ask how's your progress so far? Is it possible to finish on Thursday so we can merge it? let's meet again on Thursday in uni if you guys are free \smiley{}\\
\textbf{Student 7.3:}&Hi @Student 7.2 and @Student 7.4, hope you're well. Would like to ask the progress of your work. Thanks\\

\end{tabular}
\caption{Example of \textit{Action Redoings} - Team member Student 7.2 trying to encourage dialogue with the other team members}
    \label{fig:action_redoings_sequence_completion_example}
\end{figure}

The Pearson's chi-squared test was applied to the sequence completion dataset, except for the \textit{Silence} behaviour, which has no team members assigned to this node. Figure \ref{fig:silence_sequence_completion_example} shows an example of the \textit{Silence} behaviour, demonstrating how none of the team members responded. Table \ref{tbl:silence_complete} shows the \textit{Silence} behaviour was mainly applied to the \textit{Communication} (51.78\%, n=29) initiated behaviour, a behaviour that has team members exchanging information that sometimes does not require further discussion. For example, Student 1.4 used \textit{Communication}, stating \textit{``My ubuntu fixed itself overnight, I don't know how. Got zettlr up and running in 5 mins. what a fairy tale''}. The \textit{Silence} behaviour was rarely applied for \textit{Backup Behaviour: Seeking} (n=1, 1.79\%) and \textit{Monitoring} (n=1, 1.79\%). The Pearson's chi-squared test showed statistical differences (${\chi}^2$ $\geq$ 5 at \textit{p} $\leq$ 0.05) with \textit{Expansion - Continue} and \textit{Action Redoings}. \textit{Expansion - Continue} (W=73\%, M=27\%) and \textit{Action Redoings} (W=73\%, M=27\%) were predominately applied by the women.

\begin{figure}[b]
\centering
\begin{tabular}{p{2.5cm}p{8.5cm}}

\textbf{Student 3.1:}& Meeting Minutes for All Meetings (edited)\\
&G Suite Document $\blacktriangledown$ \\
&\includegraphics[scale=0.30]{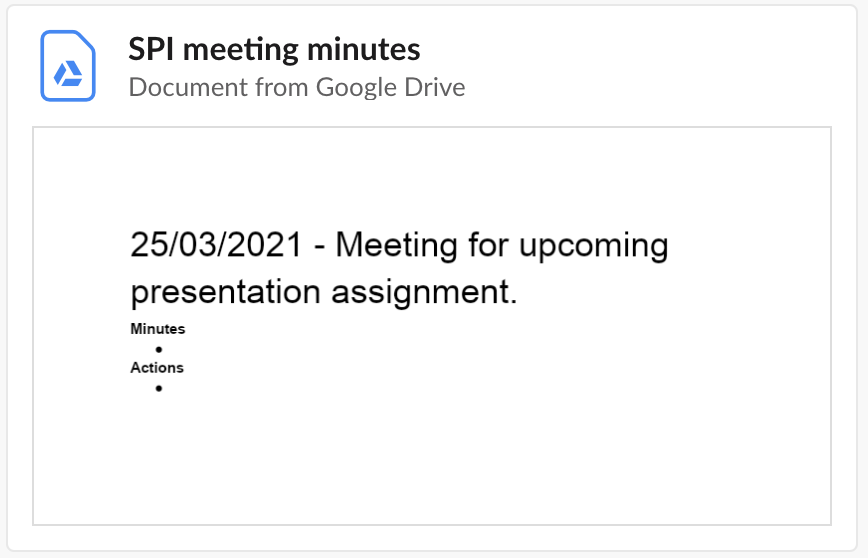} \\

\end{tabular}
\caption{Example of \textit{Silence} Sequence Completion - Shows no action from the other team members from Student 3.1's opening sequence}
    \label{fig:silence_sequence_completion_example}
\end{figure}

\textit{Delayed Replies} (${f}$=24, W=21\%, M=79\%) was used by students. Figure \ref{fig:delayedResponse} shows the distribution of \textit{Delayed Replies}, organising by teams. The figure shows the delayed replies in hours from the initiated communication. The median for the delayed replies was 13.19 hours. Most (80.77\%, n=21) of the delayed responses were performed by the men participating in the study, where Team-4 had the highest median (301.88 hours).

\begin{table}[h!]
\centering
\caption{Distribution of \textit{Silence} Sequence Completion Applied to Opening Sequences}
    \label{tbl:silence_complete}
\begin{tabular}{l|c}
\textbf{Opening Sequence} & \textbf{\textit{Silence} Applied to Sequences (\%)}  \\ \hline
Communication & 29 (51.78\%) \\
Discussing Deliverables & 7 (12.50\%) \\ 
Coordination & 6 (10.71\%)\\ 
Scheduling Meeting & 6 (10.71\%)\\ 
Team Leadership & 3 (5.36\%) \\ 
Team Orientation & 3 (5.36\%)\\
Backup Behaviour: Seeking & 1 (1.79\%) \\
Monitoring & 1 (1.79\%) \\

\end{tabular}
\end{table}

\subsubsection{Pull Request Communication} \label{subsub_pull_requests}

The final analysis if the teams' communication focused on their Pull Requests (PRs). Table \ref{pull_request_table} presents the communication frequency for the Pull Requests (PRs). The results are collated by teams, showing: 
\begin{itemize}
    \item The teams' GitHub projects,
    \item The number of PRs submitted by the teams, 
    \item The message frequencies on the GitHub and Slack platforms, and
    \item The state of the PRs.
\end{itemize}

The table organises the results by the teams' GitHub messages and includes the under- (UG) and postgraduate (PG) students in the teams. The results demonstrate communication is not an indicator of PR acceptance. The results show Team-1 having the most (n=70, 76.1\%, Accept) messages with the project maintainers, while Team-6 has the least (n=1, 5.2\%, Abandon). Table \ref{pull_request_table} shows five (62.5\%) PRs accepted, one (12.5\%) revised, and two (25.0\%) abandoned. Reasons for revised and abandoned PRs might be due to the issue being previously addressed by another open-source community member or a decision by the project maintainers not to pursue the issue. Another reason for revised and abandoned PRs might be teams running out of time to address the maintainers' feedback. Most (n=7, 87.5\%) of the teams received feedback from the maintainers, with one team, Team-3, having their PR accepted without feedback. The participants' UG and PG status did not influence the PR acceptance. For example, teams 1 (UG=4, PG=1) and 7 (UG=1, PG=4) had PRs accepted, while teams 2 (UG=1, PG=4) and 8 (UG=4, PG=0) had abandoned PRs.

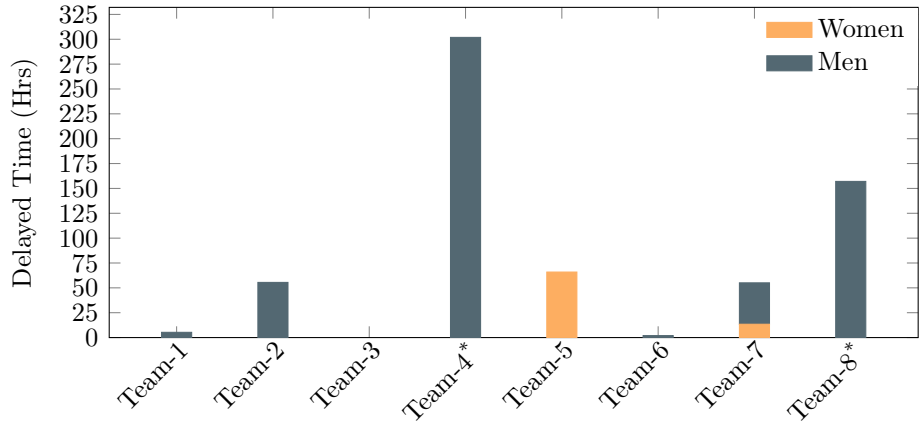
\begin{figure}[h]
\centering
    \begin{tikzpicture}
\begin{axis}[
    ybar stacked,
    legend style={
    legend columns=1,
        at={(1.00,1)},
        draw=none,
        area legend,
    },
    ytick=data,
    tick label style={font=\footnotesize},
    legend style={font=\footnotesize},
    label style={font=\footnotesize},
    ytick={0,25,50,75,100,125,150,175,200,225,250,275,300,325},
    height=170,
    width=.9\textwidth,
    bar width=4mm,
    ylabel={Delayed Time (Hrs)},
    x tick label style={rotate=45, anchor=north east, inner sep=0mm},
    xtick=data,
    xticklabels={Team-1, Team-2, Team-3, Team-4\textsuperscript{*}, Team-5, Team-6, Team-7, Team-8\textsuperscript{*}},
    ymin=0,
    area legend,
]
\addplot[female,fill=female] coordinates
{(1,0) (2,0) (3,0) (4,0) (5,65.9) (6,0) (7,13.41) (8,0)};
\addplot[male,fill=male] coordinates
{(1,5.28) (2,55.48) (3,0) (4,301.88) (5,0) (6,1.98) (7,41.63) (8,157.05)};
\legend{Women, Men}
\end{axis}  
\end{tikzpicture}
\caption{Delayed Replies Organised by Teams}
    \label{fig:delayedResponse}
\captionsetup{justification=centering}
\begin{center}
* \small{Has a non-participant team member; excluded from study}
\end{center}
\end{figure}


\begin{table}[t]
\centering
\setlength\extrarowheight{2pt}
\fontsize{10pt}{10pt}\selectfont
\caption{Communication Frequency for Pull Requests} \label{pull_request_table}
\begin{tabular}{l|c|l|c|c|c|l}
&&&\multicolumn{2}{c|}{\textbf{Communication (${f}$)}} &\\
\textbf{Team} & \textbf{UG \& PG} & \textbf{Project} & \textbf{\# PRs}&\textbf{GitHub} & \textbf{Slack} & \textbf{State}\\ \hline 
Team-1 & UG=4, PG=1 & Zettlr & 2& 70 (36.46\%) & 12 (15.00\%) & Accept \\ 
Team-7 & UG=1, PG=4 & Jabref & 1&40 (20.84\%) & 6 (7.50\%)& Accept \\ 
Team-2 & UG=1, PG=4 & Pygments & 1&34 (17.71\%) & 3 (3.75\%) & Abandon \\ 
Team-5 & UG=1, PG=3 & Zettlr & 2&27 (14.06\%) & 43 (53.75\%)& Accept \\ 
Team-4 & UG=1, PG=3 & Jabref & 1&13 (6.77\%) & 1 (1.25\%)& Revise \\
Team-3 & UG=6, PG=0 & Pygments & 1&4 (2.08\%) & 8 (10.00\%)& Accept \\
Team-6 & UG=0, PG=6 & Zettlr & 1&3 (1.56\%) & 1 (1.25\%)& Accept \\
Team-8 & UG=4, PG=0 & Pygments & 1&1 (0.52\%)& 6 (7.50\%)& Abandon \\

\end{tabular}
\end{table}

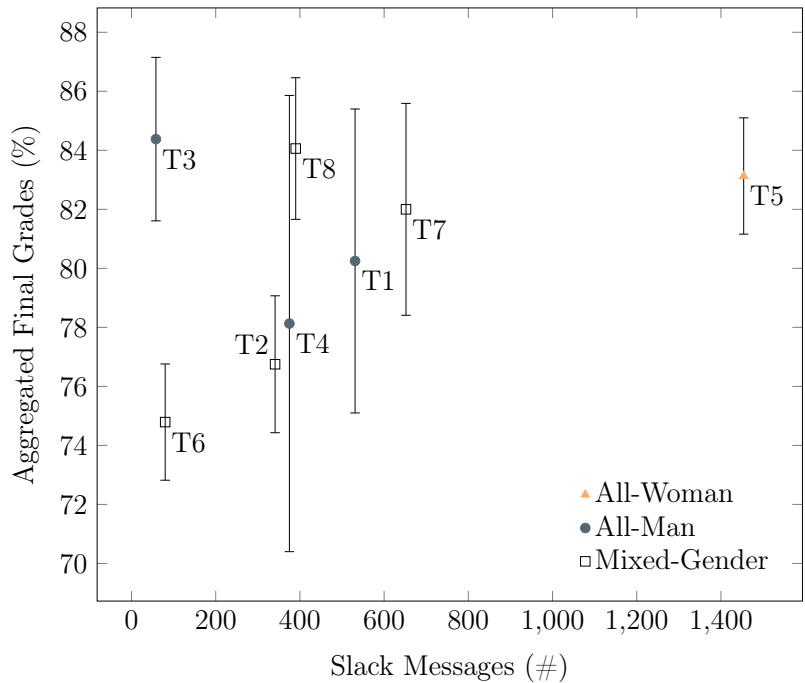
\begin{figure}[b!]
\centering
\begin{tikzpicture}[scale=.9, baseline=(current axis.outer east)]
\begin{axis}[enlargelimits=0.1,
    width=12cm,
    legend style={draw=none},
    legend pos=south east,
    xlabel={Slack Messages (\#)},
    ylabel={Aggregated Final Grades (\%)},
    ]
    \addplot[
        scatter/classes={
            f={mark=triangle*,female}, 
            m={male}, 
            b={mark=square,black}
        },
        scatter, mark=*, only marks, 
        scatter src=explicit symbolic,
        nodes near coords*={\Label},
        every node near coord/.style={anchor=\alignment},
        visualization depends on={value \thisrow{label} \as \Label}, 
        visualization depends on=\thisrow{alignment} \as \alignment,
    ]
    plot [error bars/.cd, y dir = both, y explicit]
    table [x =x, y =y, meta=class, col sep=comma, y error=error] {
        x, y, class, error, label, alignment 
        531, 80.25, m, 5.15, T1, 140 
        341, 76.75, b, 2.32, T2, -40 
        58, 84.38, m, 2.77, T3, 140
        375, 78.13, m, 7.73, T4, 140
        1454, 83.13, f, 1.97, T5, 140
        80, 74.79, b, 1.97, T6, 140
        652, 82, b, 3.59, T7, 140
        390, 84.06, b, 2.40, T8, 140
    };
    \legend{All-Woman, All-Man, Mixed-Gender}
\end{axis}
\end{tikzpicture}
\caption{Comparing Teams' Grades to Their Generated Slack Messages - Includes standard deviation (SD) of teams' grades}
    \label{fig:finalGrades}
\end{figure}

\subsection{Participants' Learning Gains} \label{section_grades}

We report on the participants’ learning gains by teams and gender, using the Slack messages generated during the teamwork. Figure \ref{fig:finalGrades} shows the comparison, presenting the teams' average grades and standard deviation (SD) for the teams' grades. The averages exclude the marks from the \textit{Team communication} assessment to remove bias. We found no relationship between the teams' average grades and the frequency of their Slack messages. For example, teams 3 ($\bar{x}$=84.38\%, SD=2.77), 8 ($\bar{x}$=84.06\%, SD=2.40), and 5 ($\bar{x}$=83.13\%, SD=7.73) had the highest average grades for the course, but the results do not demonstrate the online communication influencing the grades. Team-3 ($\bar{x}$=84.38\%, SD=2.77) generated 58 Slack messages during their collaboration but received a higher average grade than Team-5 ($\bar{x}$=83.13\%, 7.73) with 1454 messages. In addition, the results also show the participants' under- (UG) and postgraduate (PG) statuses did not influence the learning gains. For example, Team-3 is a group of undergraduates (UG=6, PG=0), while Team-5 is a blend, predominately comprised of postgraduates (UG=1, PG=3). Another example is with Team-3 ($\bar{x}$=84.38\%, SD=2.77, ${f}$=58 messages, UG=6, PG=0) and Team-6 ($\bar{x}$=74.79\%, SD=1.97, ${f}$=80 messages, UG=0, PG=6) have similar message frequencies but a 9.59\% difference in the teams’ grades. These teams had the lowest message frequencies in the study, yet they are comprised of different under- and postgraduate statuses. Team-3 comprises of six undergraduates, while Team-6 comprises of six postgraduates.

\begin{table}[t]
\centering
\caption{Marks for the Course Grade and Assessments} \label{tbl_marks_assessments}
\begin{tabular}{lc|c|c|c}
\multicolumn{2}{c|}{}&\multicolumn{3}{c}{\textbf{Average (Mean)}} \\
\multicolumn{2}{c|}{\textbf{Participants (\#)}} &\textbf{Course} & \textbf{Quiz}&\textbf{Reflection Essay} \\ \hline
\textbf{Women} & 9 & 81.67\%, SD=4.52 & 91.11\% SD=1.20& 87.22\% SD=1.11\\ \hline
\textbf{Men} & 30 & 79.88\% SD=1.86& 94.67\% SD=0.88& 80.00\% SD=1.23\\ \hline \hline
\textbf{Cohen's \textit{d\textsubscript{s}}} && 0.2929 & 0.2711 & 0.6426\\ 
\end{tabular}
\end{table}

Table \ref{tbl_marks_assessments} shows the results when evaluating the learning gains by gender. The table displays the average (mean) for the course grade and the \textit{Quiz} and \textit{Reflection essay} assessments by gender. Also included in the table is Cohen's \textit{d\textsubscript{s}}, presenting the effect size for the comparison. The table shows the women receiving higher grades (81.67\%, SD=4.52) than the men (79.88\%, SD=1.86), with higher grades (1.79\%). When comparing the course average, the overall effect size has a Cohen's \textit{d\textsubscript{s}} value of 0.2929, which is a small effect. For the individual assessments, the table shows that the men participating in the study (94.67\%, SD=0.88) received higher marks (3.56\%) than the women (91.11\%, 1.20). However, for the \textit{Reflection essay}, the women (87.22\%, SD=1.11) had higher average marks for this assessment than the men (80.00\%, SD=1.23), with 7.22\% higher marks. 

In addition, we evaluated the two individual assessments using the Wilcoxon signed-rank test (\textit{W}), to determine how the two sets of students' data and awarded marks for the \textit{Quiz} and \textit{Team communication} correlated and determine whether there is a relationship between them. The results from the Wilcoxon sign-rank test show statistical significance (\textit{W} = 59.0, \textit{p} $<$ .0001) for the \textit{Quiz} assessment when correlated to the \textit{Team communication}, but no statistical significance (\textit{W} = 163.5, \textit{p} = .1556) when correlating the \textit{Reflection essay} with \textit{Team communication}. The effect size when comparing the average of the \textit{Quiz} by the women (91.11\%, SD=1.20) and the men (94.67\%, SD=0.88) have a Cohen's \textit{d\textsubscript{s}} value of 0.2711, a small effect size, while the \textit{Reflection essay} had a Cohen's \textit{d\textsubscript{s}} value of 0.6426. a medium effect size.

\section{Discussion}

This paper presents a study examining Software Engineering students' communication skills while working in teams to contribute to a large open-source software project. Section \ref{section_results} presented the results. In this section, we discuss the findings. Section \ref{subsection_threats} presents the limitations and threats to validity. Sections \ref{subsection_rq1}, \ref{subsection_rq2}, and \ref{subsection_rq3} discuss the study's findings within the context of the three research questions. We conclude our paper with Section \ref{subsection_future}, discussing future research opportunities from our findings.

\subsection{Threats to Validity} \label{subsection_threats}

There are limitations and contextual variables in this study. Volunteer bias might have influenced participants' communications, potentially showing them on their ``best behaviour'' or exhibiting performative behaviour for higher marks on the \textit{Team Communication} assessment. As previously mentioned in Section \ref{survey_conversation_analysis_opening_sequences}, our ethics approval constrained student communications to the teams' Slack channels. It is possible students communicated through private Slack messages or another communication platform. However, our study collected and analysed 3881 Slack messages, providing assurance the teams used Slack while working on their projects. We were also limited, due to the ethics requirements, analysing the participants' grades to protect their identity. Another limitation, due to the ethics requirements, is our inability to capture teams' communication outside of Slack, such as the teams' face-to-face meetings and with the project maintainers. The face-to-face meetings might have promoted gender-related behaviours influencing teams' Slack communication and team members' self-efficacy during the collaboration process. In addition, even though the lecturer encouraged students to use Slack for all team communications for assessment, we cannot say for certain whether teams or individuals within the teams communicated on other asynchronous collaboration platforms, but as previously mentioned, we collected 3881 Slack messages, providing assurance the students used the team Slack channels.

A threat to validity is our inability to control the project maintainers' response times or interpersonal interactions. With the teams working on different open-source projects, the different project maintainers' responses could have influenced the participants' behaviours. Though we report on the acceptance of the teams' PRs, multiple factors influence the acceptance rate, such as the project maintainers' commitments to other tasks or waning interest in the feature. Another limitation is the reporting from one all-woman team, which is an unequal representation compared to the all-man (n=3) and mixed-gender (n=4) teams. In addition, our study did not include a mixed-gender team predominately made up of women to identify the teamwork behaviours that emerge from this team composition. While we cannot claim that our findings generalise to other teamwork, the projects were designed to represent real-world software development settings as best as possible in a university setting by asking students to contribute to large open-source projects.

Lastly, we also recognise our study produced small effect sizes, potentially due to the small sample size limiting us from making significant correlations between the collected data and the under-represented group of women. However, though our sample size is small, trends that emerged from the collected data warrant further research and could help practitioners realise the different behaviours and communications that occur with team Software Engineering projects. Though some \citep{vecchio:2002} argue sample effect sizes are irrelevant, others \citep{rosenthal:1994} view small effect sizes as having practical implications and relevance to real-life settings. In our case, with the women participating in the study, the group work may have significantly influenced their perceptions of a career in computing. Inequitable and negative experiences could affect their decisions to continue their studies, and practitioners could include interventions that mitigate these experiences. Computing education researchers often remove the demographics of marginalised groups for analysis, where exclusion reasons include small data sets that lead to a lack of representation \citep{oleson:2022}. Through our contribution discussing outcomes on women in computing, future computing education researchers can conduct replication studies, performing further interactions so that this area of research can get a global perspective of students' communication during group work.

\subsection{RQ1: How does gender influence students' initiation of communication within teams?} \label{subsection_rq1}

In Section \ref{section_results}, the post-survey results show both the women and men participating in the study had similar positive collaboration experiences during the study, but their communication behaviours varied. For example, Figures \ref{fig:teamcompositions} and \ref{fig:genderMessages} demonstrate that the women were more communicative, especially within the all-woman team. Prior research \citep{rogelberg:1996} has shown teams with women were more productive. A higher communication rate could contribute towards the team's productivity, but within our study, the women were more communicative within certain conditions. Within the all-woman team, team members showed a higher rate of seeking peer help. A potential reason for their help-seeking is \textit{peer parity}, which occurs ``when an individual can identify with at least one other peer when interacting in a community'' \citep[p.~1]{ford:2017}. Peer parity might have been present within Team-7 (Women=2, Men=3), encouraging higher engagement by the two women in the team. Within the mixed-gender teams with a single woman (Teams 2 and 6), the absence of peer parity might have contributed to their low team communications. These participants might have felt isolated within their team. Feelings of isolation can be a demotivating factor that impedes learning \citep{sankar:2015}. Our findings differ from prior research \citep{rogelberg:1996} of mix-gendered teams in the industry, where teams with one woman were more productive than the all-man teams participating in the study. The lack of help-seeking behaviour in the men may be due to the appearance they want to project to their peers, known as \textit{effortless achievers} \citep{dweck:2002}. Effortless achievers want to appear naturally intelligent, and a trait predominately observed in younger males \citep{schneider:2005}. They are encouraged by the competitive nature of friendships and want to appear tougher without the need to study, a perceived feminine trait.

We observed the \textit{Feedback} initiated communication behaviour was not applied by the teams during the study. \textit{Feedback} occurs when a team member gives, seeks, and receives information on their peers' contributions. The absence of this behaviour might be due to the feedback coming from the project maintainers. The team members relayed the maintainers' feedback as \textit{Coordination} behaviours. For example, \textit{``we got the reply from the repo maintainer''} and \textit{``our code has been merged!''}. Analysing external factors, such as the communication with the GitHub maintainers and face-to-face meetings, might help us better understand the absence of the \textit{Feedback} behaviour.

Though team roles were not defined in this study, teamwork behaviours did materialise, as was previously observed \citep{strijbos:2005} within non-role teams. Certain teamwork behaviours, \textit{Team Leadership}, \textit{Coordination}, and \textit{Monitoring}, were primarily initiated by the women. The behaviours can require more effort from the speaker. For example, with the \textit{Monitoring} behaviour, the speaker has to assess the project status, which involves evaluating its progress \citep{martens:2019}. If the project falls short of expectations, the student initiating the \textit{Monitoring} behaviour will try to implement corrective measures to realign the project's trajectory. 

These teamwork behaviours, \textit{Team Leadership}, \textit{Coordination}, and \textit{Monitoring}, could be perceived as \textit{pink tasks}; these are tasks that need ``to be done on time and to a high standard, but where there is little substantive development or increased visibility for the person undertaking or assigned the tasks'' \citep[p.~3]{brough:2011}. We surmise the initiated behaviours within Team-7 might be the result of \textit{perceived feminine competencies}, a gender stereotype characterising women with better workplace relationship and communication competencies \citep{trauth:2012}; and as a result, Team-7 might have assumed that the women on the team would take on these roles. More research is needed to better understand the formation of teamwork behaviours. In Section \ref{section_team_collaboration}, we reported that teams with one woman did not initiate communication as much as Team-7 with two women in the mixed-gender team. Our findings support Lina Battestilli's perspective in a SIGCSE `18 panel discussion \textit{``Best Practices in Academic to Remedy Gender Bias in Tech''}, where she suggested having at least two women in a team ``so they don't feel as a minority and thus can feel comfortable speaking up'' \citep[p.~672]{wolz:2018}.

\subsection{RQ2: How does gender influence students' responses to the initiated communications?} \label{subsection_rq2}

Section \ref{section_results} presented how the teams involved in the study responded to initiated communications. Analysing the responses provides insight into how women and men communicate with each other during the collaborative software development process. Through this study, our observations help raise awareness among practitioners on how their students may communicate during collaborative activities. Table \ref{tbl:sequence_completion} shows how the women and men participating in the study responded differently to the initiated communications. For example, the table shows that the women in the study tried to complete conversations through the \textit{Expansion - Continue} and \textit{Action Redoings} behaviours. The \textit{Expansion - Continue} behaviour is when a speaker continues the sequence with the same course of action; we observed the women using this behaviour to collect more information to complete the action on the project. The \textit{Action Redoings} behaviour is when the speaker can position themselves after the course action, which indicates to the listeners that they are still responding, providing an opportunity for the next turn in the conversation to be taken by another participant \citep{hoey:2017}. We observed women using this behaviour to encourage the team to work on a task or follow up with a peer on one. We observed women using these behaviours to engage organisational and planning skills necessary to complete the software projects. The \textit{Action Redoings} shown in Figure \ref{fig:action_redoings_sequence_completion_example} is self-repetition that repeats the original sentiment without adding new information \citep{crible:2020}. However, self-repetition avoids communication breakdowns and is an indirect problem-solving strategy, which requires efforts to help keep the communication going \citep{dornyei:1995}. Some expressly essentialist views \citep{deakin:1984} consider organisational skills are predominately applied by women, perceiving they can address the communication gaps in software development due to men lacking these communication and management skills in industry \cite{whitehouse:2007}. Though these sentiments are ``traditional patterns of gender segmentation'' \citep[p.~88]{whitehouse:2007}, our results show that the women applying \textit{Expansion - Continue} and \textit{Action Redoings} behaviours requiring more effort from the women in our study, demonstrating inequitable behaviours between the women and men in the study.

On the other hand, the men in our study used more of the \textit{Post-Sequence Transition} and \textit{Delayed Replies} behaviours. The \textit{Post-Sequence Transition} behaviour is when the team considers the lapse in communication as an opportunity to shift the conversation topic to something else. It is unclear from the teams' Slack channels how they addressed tasks that did not receive peer attention. It is possible that the person initiated the communication independently addressing the task, potentially generating more work for the student than their peers or the team left the task unfinished. More work is required to understand how the team addressed these tasks. The \textit{Delayed Replies} behaviour is when another member of the team responds to the initiated communication.

Our findings show that the men in the study used more delayed replies than the women. The delayed replies might be a form of procrastination, where the learner tends to postpone or delay a series of tasks \citep{milgram:1991}. Prior research has focused on procrastination through the lens of gender, producing mixed findings. One procrastination study by \citet{rodarte:2008} concluded that the women in that study frequently procrastinated, while a study by \citet{balkis:2009} showed higher procrastination rates among pre-service educators than men. Another study \citep{ozer:2011} with high school and university students in Turkey showed similar procrastination rates across genders. Within the computing discipline, Masters Computer Science students in Northeast Argentina showed no differences between the women and men in their study \citep{irrazabal:2017}. Instead of procrastination, another possibility for the delayed replies might be students' varied hours working on their projects. The communications on the teams' Slack channels seem to occur at all hours of the day, which is different to the work behaviours observed in professional software development with defined hours. The lack of defined work hours for the student teams could explain the lengthy response time (median 13.19 hours) in our teams. At the same time, the majority (61.9\%) of professional software developers' communications received an answer within an hour \citep{elmezouar:2021}. Additional work is required to understand the delayed replies and how educators can help through interventions and strategies.

Our results show the participants rarely (1.04\%, n=4) use the \textit{Turn-Exiting} behaviour, a behaviour in which a participant abandons the communication sequence. \textit{Turn-Exiting} occurs when a person wants to exit a conversation \citep{hoey:2017}. Within oral communication, speakers use turn-exiting to disengage from the conversation \citep{rossano:2012}, to acknowledge a previous speaker's turn \citep{schegloff:1982}, and to ``exit the sequence on an amicable note'' \citep[p.~148]{schegloff:2007}. Within asynchronous, online communications, the application of turn-exiting might be perceived as unnecessary and instead applied the \textit{Silence} behaviour, but more work is required to confirm our theory. 

When analysing how the women and men completed communications, we observe inequities in how the women and men addressed the communication. To engage certain behaviours, such as \textit{Expansion - Continue}, requires additional effort from the speaker to contribute to the conversation. Likewise, the \textit{Action Redoings} behaviour requires skills to encourage others to engage in a conversation that the others previously ignored. A potential reason for peers not responding and speakers engaging in \textit{Action Redoings} may be due to communication apprehension, a behaviour previously observed by students in collaborative learning environments \cite{hunt:2004}. In this research, \citet{hunt:2004} did not specify a particular gender that predominately uses communication apprehension. However, more work is required to understand why students, especially women, apply these behaviours.

\subsection{RQ3: How do the communication behaviours between women and men in collaborative team settings impact learning gains?} \label{subsection_rq3}

To answer this research question, we examined participants' academic performance to determine whether team collaboration positively influenced their learning gains. The evaluation examined learning gains from the teams' and genders' perspectives, comparing them to the frequency of communication in the team's Slack messages and the communication behaviours they used on the channels. From the team perspective, the results showed the frequency of students' online team communication and their communication behaviours did not influence their learning gains. While conducting this evaluation, we considered the students' under- (UG) and postgraduate (PG) status but did not see the status having an influence on the students' final grades for the course. More work is required to explore the difference between UG and PG students and how they might apply communication differently. Our results show participants' under- and postgraduate status did not factor into teams' performance. In addition, our evaluation of the grades and communication through the lens of gender found similar results to the team comparison: the frequency and communication behaviours had no influence on the course grades awarded to the women and men participating in the study. Though another study \citep{pepe:2012} identified a correlation between learning gains and study skills, our study was conducted with in-person learning using individual assessments. More research is required to determine whether the correlation between learning gains and study skills exists in online learning environments using collaborative learning activities like our study. 

When we compared the communications with the marks of individual assessments, we observed a statistical significance in comparing the marks awarded for team communications and for the \textit{Quiz} summative assessment. The statistical significance (\textit{W} = 59.0, \textit{p} $<$ .0001) in the Wilcoxon signed-rank test (\textit{W} between the \textit{Quiz} and \textit{Team communication} data sets may show how the communication influenced the students' understanding of the Software Engineering concepts taught in the course. In the course, the \textit{Quiz} was administered in a computer-based testing environment to reinforce students' understanding of contributions made to open-source projects. The higher engagement through online team communication, measured by the \textit{Team communication} dataset, might have given students with higher communication rate a more in-depth understanding of the software development process. It has been previously shown \citep{pekrun:2012}that  students who are more engaged during the learning process tend to have higher course achievements, which may explain our findings comparing the marks from the \textit{Quiz} and the marks awarded for the \textit{Team communication}. 

However, the statistical analysis, comparing the \textit{Reflection essay} assessment with the \textit{Team communication}, did not produce a similar correlation as the \textit{Quiz}. We did observe the women participating in the study receiving higher (7.22\%) marks in the \textit{Reflection essay} than the men. A potential reason for the higher marks may be the type of communication behaviours the women applied with their teams. As previously mentioned when answering the first and second research questions, Sections \ref{subsection_rq1} and \ref{subsection_rq2}, the women applied behaviours, such as \textit{Monitoring} and \textit{Action Redoings}, that require more  consideration. The two-page reflection essay had students reflect on their impressions before and after contributing to the open-source project. As previously discussed in Sections \ref{subsection_rq1} and \ref{subsection_rq2}, women used communication behaviours that required the speakers to measure the project and evaluate its progress \citep{martens:2019}. Monitoring also includes the individual employing corrective measures to realign in the project's trajectory if it falls short of its expectations. These activities help the speaker get involved in the project's tasks and help them observe what is required to complete the project. Getting involved firsthand in the project's progress might have helped the women in the study understand what is involved to complete an open-source project. The communication behaviours men used in the study, such as \textit{Post-Sequence Transition}, did not require the same effort as the communication behaviours applied by the women. \textit{Post-Sequence Transition}, the speaker considers the lapse an opportunity to shift the sequence's topic to something else, which might require little effort from them to use in the team communications. These communication behaviours used by the men in the study may give them a different depth of knowledge of the project than the women contributing to the conversations. More research is required to confirm the type of communication behaviours that encourage more in-depth knowledge of collaborative projects and how practitioners can encourage all students, regardless of gender, to apply a variety of communication behaviours that promote equitable roles and responsibilities in the project, along with a deeper understanding of what is required to complete it. In addition, for the \textit{Reflection essay}, the medium effect size (Cohen's \textit{d\textsubscript{s}} = 0.6426) for this assessment warrants more research to understand how the knowledge gained from team communication can be transferred to self-reflection learning activities.

In addition, our results show communication was not an indicator of the acceptance of teams' Pull Requests (PRs). Our results (See Table \ref{pull_request_table}) demonstrate a blend of communication rates for the accepted PRs, ranging from low to high team communications. The results also indicate the participants' under- and postgraduate status influence the accepted and abandoned PRs; therefore, we cannot determine how team communication affects the acceptance of Pull Requests. As stated in Section \ref{subsection_threats}, outside factors, such as project maintainers' workload, might influence PR acceptance. Still, more work is necessary to understand how these external factors influence PRs to teach students to promote higher acceptance of PRs.

\subsection{Conclusion and Future Work} \label{subsection_future}

In this paper, we conduct a study examining the Software Engineering (SE) student communications while collaborating on an open-source software project. The study focuses on interpersonal communications through the lens of gender since gender is a notable social category we wanted to use as a basis for future research. Future Software Engineering Education research can re-examine or replicate our work to examine and consider how other social categorisations and intersectionality influence students' communications. The results from this study observed different communication behaviours in SE student teams applied by women and men in the study, including how teams responded to these communications. To the best of our knowledge, our study is the first attempt at examining communication exchange within SE student teams to observe how the different communication behaviours women and men use during the collaboration of large open-source Software Engineering projects. 

We used gender analysis to examine differences in teamwork behaviours between the women and men studying Software Engineering while contributing to a large open-source Software Engineering project. The course included under- and postgraduate students whose graduate status did not influence the findings, along with prior teamwork experiences, but explains how future Software Engineering Education research can build on our findings. Without defining roles for the teams, the women in the study initiated more behaviours related to leadership, coordination, and project monitoring to help their teams complete the project, which can require additional  consideration from the speaker over other communication behaviours. We found the all-woman team acknowledged their peers' communications and responded to knowledge-sharing and help-seeking requests. In contrast, the all-man and mixed-gender teams participating in our study exhibited infrequent help-seeking behaviour. Though these teamwork behaviours were championed by men in the all-man teams, as demonstrated in Team-4, the application of these behaviours was infrequent compared to the application of these behaviours by the women. The outcomes from this study could assist researchers and practitioners in foreseeing obstacles students, and perhaps extend to software professionals, might encounter collaborating on software projects using communication platforms in comparable learning environments to our work.

Reactions to communications also showed different behaviours from the women and men participating in our study. The men in the study took longer to respond to communications. More research is necessary to identify factors contributing to the acceptance of Pull Requests and higher learning gains. Our findings raise future Software Engineering Education research opportunities to investigate further the behaviours observed in this study and evaluate the communication considering other social categorisations. Future research can also evaluate interventions and strategies that promote change in inequitable communications during SE student group work. For example, additional research can further evaluate mixed-gender team dynamics to determine whether feelings of isolation or gendered factors influence the teamwork behaviours of teams containing one woman. In addition, future work can investigate diversity and inclusion training to promote more equitable communication during group work. The student training could encourage more discussions among the group and may increase their understanding of the software development process.

\bibliographystyle{elsarticle-harv}\biboptions{authoryear}

\end{sloppy}
\end{document}